\documentclass[10pt,twocolumn,aps,prl,amsmath,superscriptaddress,notitlepage,longbibliography,nofootinbib]{revtex4-1}

\usepackage{amssymb,amsmath,amsthm,mathtools,bm,dsfont,braket,graphicx,mathrsfs}
\usepackage[utf8]{inputenc}
\usepackage[final,protrusion=true,expansion=true]{microtype}
\usepackage{etoolbox}
\usepackage{xcolor}
\colorlet{MAGENTA}{magenta}
\usepackage{booktabs}
\usepackage{multirow}
\usepackage{float}
\usepackage{algorithm}
\usepackage{algpseudocode}
\usepackage{array}
\usepackage{capt-of}
\usepackage{xparse}
\usepackage{adjustbox}
\usepackage{tikz}
\usepackage{quantikz}
\usepackage{pgfplots}
\pgfplotsset{compat=1.18}
\usepgfplotslibrary{statistics,groupplots}
\usepackage{hyperref}
\newif\ifarxivSIbibliography
\usepackage{empheq}
\hypersetup{colorlinks=true,citecolor=blue,linkcolor=blue,urlcolor=blue}
\definecolor{iceBlue}{HTML}{4B5CC4}
\definecolor{headingRed}{HTML}{9B3D4B}
\definecolor{bostonBlue}{HTML}{0072B2}
\definecolor{kingstonOrange}{HTML}{D55E00}
\definecolor{tiltA}{HTML}{5689BA}   %
\definecolor{tiltB}{HTML}{3C6E9E}   %
\definecolor{tiltC}{HTML}{173F5F}   %
\definecolor{rawA}{HTML}{D66773}    %
\definecolor{rawB}{HTML}{B63A4D}    %
\definecolor{rawC}{HTML}{8E1B2F}    %
\definecolor{floquetBlock}{HTML}{A8DCE3}
\definecolor{floquetExchange}{HTML}{F2D2B2}
\definecolor{floquetDiagonal}{HTML}{CAE1D1}
\hypersetup{colorlinks=true,linkcolor=iceBlue,citecolor=iceBlue,urlcolor=iceBlue}
\newcommand{\maintextheading}[1]{\noindent\textcolor{headingRed}{\textit{#1.---}}}
\graphicspath{{./}}

\newcommand{\compactmathspacing}{%
  \setlength{\abovedisplayskip}{4.5pt plus 1pt minus 1pt}%
  \setlength{\belowdisplayskip}{4.5pt plus 1pt minus 1pt}%
  \setlength{\abovedisplayshortskip}{1.5pt plus 1pt}%
  \setlength{\belowdisplayshortskip}{3pt plus 1pt minus 1pt}%
  \setlength{\jot}{1.25pt}%
  \setlength{\mathsurround}{0pt}%
}
\AtBeginDocument{%
  \apptocmd{\normalsize}{\compactmathspacing}{}{}%
  \apptocmd{\small}{\compactmathspacing}{}{}%
  \compactmathspacing
  \setlength{\parskip}{0pt}%
  \setlength{\emergencystretch}{0.6em}%
  \clubpenalty=5000
  \widowpenalty=5000
}

\AtBeginDocument{%
  \setlength{\heavyrulewidth}{0.8pt}%
  \setlength{\lightrulewidth}{0.5pt}%
  \setlength{\cmidrulewidth}{0.3pt}%
  \setlength{\aboverulesep}{2.0pt}%
  \setlength{\belowrulesep}{3.25pt}%
}

\makeatletter
\def\title@column@grid#1{%
  \minipagefootnote@init
  \onecolumngrid
  \begingroup
    \let\@footnotetext\frontmatter@footnotetext
    \let\set@footnotewidth\set@footnotewidth@one
    \ltx@no@footnote
    #1%
  \endgroup
  \minipagefootnote@foot
  \twocolumngrid
}
\makeatother

\makeatletter

\newcommand{\startsupplementaryfloatlists}{%
  \def\ext@figure{slof}%
  \def\ext@table{slot}%
}

\newcommand{\listsupplementaryfigures}{%
  \begingroup
    \let\toc@post\@empty
    \print@toc{slof}%
  \endgroup
}
\newcommand{\listsupplementarytables}{%
  \begingroup
    \let\toc@post\@empty
    \print@toc{slot}%
  \endgroup
}
\makeatother

\newcommand{\compactfloatspacing}{%
  \setlength{\textfloatsep}{11pt}%
  \setlength{\dbltextfloatsep}{8pt}%
  \setlength{\floatsep}{6pt}%
  \setlength{\dblfloatsep}{6pt}%
  \setlength{\intextsep}{6pt}%
  \setlength{\abovecaptionskip}{5pt}%
  \setlength{\belowcaptionskip}{0pt}%
}

\newcommand{\Tr}{\operatorname{Tr}}
\newcommand{\ii}{\mathrm{i}}
\newcommand{\dd}{\mathrm{d}}
\newcommand{\cP}{\mathcal{P}}

\newcommand{\Ba}{\mathcal B_a}

\newcommand{\Ifour}{\mathcal I_4}

\newcommand{\insertdevicepropertiespdf}{%
  \includegraphics[width=0.74\textwidth]{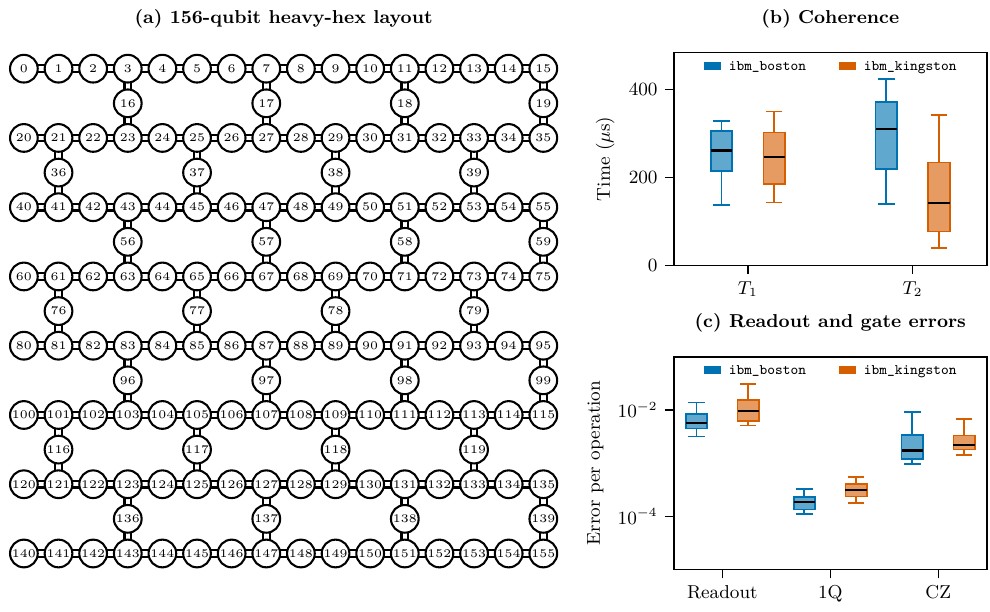}}

\begin{document}

\title{Observing the magic Mpemba effect in localized dynamics\\
on a digital quantum computer}

\author{Han-Ze Li}
\thanks{These two authors contributed equally to this work.}
\affiliation{Department of Physics, National University of Singapore, Singapore 117551}
\affiliation{Institute for Quantum Science and Technology, Shanghai University, Shanghai 200444, China}

\author{Xianquan Yan}
\thanks{These two authors contributed equally to this work.}
\affiliation{Department of Physics, National University of Singapore, Singapore 117551}
\affiliation{Department of Computer Science, National University of Singapore, Singapore 117417}

\author{Yi-Rui Zhang}
\affiliation{Institute for Quantum Science and Technology, Shanghai University, Shanghai 200444, China}

\author{Jian-Xin Zhong}
\email{jxzhong@shu.edu.cn}
\affiliation{Institute for Quantum Science and Technology, Shanghai University, Shanghai 200444, China}

\author{Shuo Liu}
\email{sl6097@princeton.edu}
\affiliation{Department of Physics, Princeton University, Princeton, New Jersey 08544, USA}

\author{Ching Hua Lee}
\email{phylch@nus.edu.sg}
\affiliation{Department of Physics, National University of Singapore, Singapore 117551}

\begin{abstract}
The Mpemba effect challenges the intuition that states closer to equilibrium must relax faster. We ask whether an analogous magic-ordering reversal can occur in the generation of quantum magic (nonstabilizerness), a key resource for universal quantum computation: can a state with less initial magic overtake one with more and reach its asymptotic value sooner? Here, we uncover interaction-induced dephasing as a distinct mechanism for this magic Mpemba effect, requiring neither transport nor conventional thermalization. For tilted product states in an interacting $\ell$-bit model, a random-phase analysis predicts a reversal of the magic ordering: states with less initial magic attain higher saturated magic, independent of their spatial pattern. Exact simulations of the $\ell$-bit model and localized spin chains further show that these states also saturate earlier. We observe both features on IBM superconducting quantum processors. An effective $\ell$-bit implementation accesses long dephasing times at fixed circuit depth by encoding time in gate angles, while a microscopic spin model exhibits the reversal universally across initial patterns in the localized regime but only selectively in the ergodic regime. These results establish interaction-induced dephasing as a route to anomalous quantum-resource relaxation beyond transport-driven thermalizing dynamics.
\end{abstract}

\maketitle
\raggedbottom

% Compact spacing for the main Letter only; retain the 10-pt font and class margins.
\newlength{\savedAboveDisplay}\setlength{\savedAboveDisplay}{\abovedisplayskip}
\newlength{\savedBelowDisplay}\setlength{\savedBelowDisplay}{\belowdisplayskip}
\newlength{\savedAboveShortDisplay}\setlength{\savedAboveShortDisplay}{\abovedisplayshortskip}
\newlength{\savedBelowShortDisplay}\setlength{\savedBelowShortDisplay}{\belowdisplayshortskip}
\newlength{\savedJot}\setlength{\savedJot}{\jot}
\newlength{\savedTextFloatSep}\setlength{\savedTextFloatSep}{\textfloatsep}
\newlength{\savedDblTextFloatSep}\setlength{\savedDblTextFloatSep}{\dbltextfloatsep}
\newlength{\savedInTextSep}\setlength{\savedInTextSep}{\intextsep}
\newlength{\savedFloatSep}\setlength{\savedFloatSep}{\floatsep}
\newlength{\savedDblFloatSep}\setlength{\savedDblFloatSep}{\dblfloatsep}
\compactmathspacing

% Keep caption spacing compact within the main Letter.
\newlength{\mainSavedAboveCaption}\setlength{\mainSavedAboveCaption}{\abovecaptionskip}
\newlength{\mainSavedBelowCaption}\setlength{\mainSavedBelowCaption}{\belowcaptionskip}
\compactfloatspacing

\maintextheading{Introduction} The Mpemba effect, in which an initially hotter system can cool faster than a colder one~\cite{MpembaOsborne1969,Kell1969Freezing,Auerbach1995Supercooling,Jeng2006Mpemba}, has become a paradigm of anomalous relaxation far from equilibrium~\cite{LuRaz2017Markovian,KlichRazHirschbergVucelja2019,HJXMeanField,HJXFirstOrderPT,SpeedupsInNonequilibriumThermalRelaxation,Lasanta2017Granular,BaityJesi2019SpinGlasses,KumarBechhoefer2020Cooling,HJXMeanfield2,HJXOrderParameter}. Quantum Mpemba effects have since been studied both in relaxation toward stationary states~\cite{Ares2025QuantumMpemba,QuantumMpembaEffectsFromSymmetryPerspectives,TheQuantumMpembaEffectInClosedSystemsFromTheoryToExperiment,Ares2025SimplerProbe,QuenchingFromSuperfluidToFreeBosonsIn,Panduranghan2026ClosedThermalisation} and in dynamical symmetry restoration~\cite{NonEquilibriumEntanglementAsymmetryForDiscreteGroups,Russotto2026QSSEPMpemba}, where a state that initially breaks a symmetry more strongly can restore it locally faster~\cite{EntanglementAsymmetryAsAProbeOfSymmetry,Rylands2024,LiuZhangYinZhang2024,EntanglementAsymmetryAndQuantumMpembaEffectIn,MultipleCrossingsDuringDynamicalSymmetryRestorationAnd,EntanglementAsymmetryAndQuantumMpembaEffectInX,DynamicalSymmetryRestorationInTheHeisenbergSpin,TurkeshiCalabreseDeLuca2025,Foligno2025ChargedDualUnitary,YamashikaAres2026LongRange,Yamashika2026SymmetricCliffordMpemba,Vescovo2026KramersWannierMpemba,Das2026FloquetMpemba,QuantumMpembaEffectInFreeFermionicMixed,TranslationSymmetryRestorationUnderRandomUnitaryDynamics,EntanglementAsymmetryDynamicsInRandomQuantumCircuits,TuningTheQuantumMpembaEffectInAn,Liu2025MBLQME,StarkManyBodyLocalizationInducedQuantumMpemba,QuantumMpembaEffectInLongRangedU}. Anomalous relaxation has been observed in dissipative trapped-ion systems~\cite{AharonyShapira2024} and symmetry restoration in trapped-ion and superconducting simulators~\cite{Joshi2024ObservingQME,xu2025observationmodulationquantummpemba}. A resource-theoretic formulation unifies these phenomena as anomalous resource depletion~\cite{ResourceTheoreticalUnificationOfMpembaEffects}: an initially more resourceful state loses its resource faster under the same free dynamics. We refer to this as the ``cooling-side'' picture. Its complement is resource generation: can an initially less resourceful state overtake a more resourceful one under the same resource-generating dynamics? Such a ``heating-side'' Mpemba effect is directly relevant to resource-state preparation and has recently been explored for entanglement~\cite{bao2026entanglementmpembaeffect,benjadi2026exponentialspeedupentanglementgeneration}.

Magic, or nonstabilizerness, quantifies departure from stabilizer states and is a key resource for quantum computation beyond classically simulable stabilizer operations~\cite{gottesman1998heisenbergrepresentationquantumcomputers,PhysRevA.70.052328,BravyiKitaev2005,Veitch2012Negative,Veitch2014Resource,ContextualitySuppliesTheMagicForQuantumComputation,HowardCampbell2017,BermejoVega2017,QuantifyingQuantumSpeedupsImprovedClassicalSimulationFrom,MagicStateDistillationWithLowOverhead,qian2025quantum}. Recent work has uncovered universal aspects of its growth and spreading~\cite{MagicSpreadingInRandomQuantumCircuits,tirrito2025universalspreadingnonstabilizernessquantum}, while its many-body dynamics~\cite{LiuWinter2022ManyBodyMagic,White2021MagicalCFT,QuantumMagicDynamicsInRandomCircuits,MagicSpreadingInRandomQuantumCircuits,tirrito2025anticoncentrationnonstabilizernessspreadingergodic,aditya2025growthspreadingquantumresources,Magni2025quantumcomplexity,iannotti2026nonstabilizernessu1symmetrychaotic,liu2026nonlocal,sierant2026exactquantificationnonlocalmagic,v3f2-ldr1,xiao2026diffusivedynamicsnonstabilizerness,Rattacaso2023Quench,Turkeshi2025PauliSpectrum,Turkeshi2026GaussianMagic,Li2026NonlocalMagicMBL,Turkeshi2023MultifractalFlatness,Yang2017EntanglementComplexity} and connections to other quantum resources~\cite{zhang2026revealingentanglementgrowthmechanismsmagic,zhang2026entanglementgrowthtransportschmidt,aditya2026equivalencequantumresourcesergodic,liu2026entanglementantiflatnessnonlocalnonstabilizerness,Tirrito2024EntanglementFlatness,Hou2026MagicInjection,Aditya2026NonlocalMagicSpreading,Sierant2026SpectralGeometry} have been widely investigated. No cooling-side Mpemba reversal was found for the local magic of tilted product states under random Clifford dynamics~\cite{aditya2025mpembaeffectsquantumcomplexity}, whereas a heating-side \emph{magic Mpemba effect} (MME) was recently identified in chaotic symmetry-conserving circuits~\cite{Xiao2026NonstabilizernessMpemba}.

This raises a basic question: does the MME rely on transport and chaotic relaxation, or can it persist when both are suppressed? In chaotic circuits, the effect depends on the initial-state family~\cite{Xiao2026NonstabilizernessMpemba}, and related studies have linked quantum-resource dynamics to transport~\cite{tirrito2025universalspreadingnonstabilizernessquantum,xiao2026diffusivedynamicsnonstabilizerness,Zhang2026EntanglementEmbezzlement,Huang2025FeedbackEntanglement}. Localized systems~\cite{Gornyi2005Localization,Basko2006Localization,Nandkishore2015MBLReview,Abanin2019MBLReview,Sierant2025MBLReview}, by contrast, suppress transport~\cite{Anderson1958Localization} and evade conventional thermalization~\cite{Serbyn2013LocalConservation,Huse2014Phenomenology,Serbyn2013UniversalDephasing,Serbyn2014QuantumQuenches}. Although interaction-induced dephasing is known to generate magic in localized dynamics~\cite{xfp5-hhs4,195d-r5j3}, establishing an MME requires more: states with less initial magic must overtake more resourceful states and reach their respective asymptotic values sooner.

A complementary challenge concerns hardware realization. Programmable quantum processors provide controlled state preparation, interactions, and measurements~\cite{Feynman1982QuantumSimulation,Lloyd1996UniversalSimulators,Georgescu2014QuantumSimulation,Lanyon2011DigitalSimulation,Barends2015DigitalFermions,Koh2026EdgeClusterBursts,Koh2024HigherOrder,Kim2023QuantumUtility,Shen2026CondensedMatterReview,Li2026PracticalErrorReview,Liu2023MBLVQE,Shen2026TopologicalOrder,xu2026fractional,Yang2026ProbingEntanglementSymmetries}, and magic has been measured in circuit-prepared states~\cite{Oliviero2022MeasuringMagic,ScalableMeasuresOfMagicResourceForQuantum}. However, resolving an MME in localized dynamics requires accurate higher-order Pauli moments~\cite{Leone2022SRE,StabilizerEntropiesAndNonstabilizernessMonotones,Haug2024StabilizerEntropies} over the long timescales of dephasing, while gate errors can themselves generate spurious magic~\cite{Oliviero2022MeasuringMagic}.

\begin{figure*}[!t]
  \centering
  \includegraphics[width=0.88\textwidth]{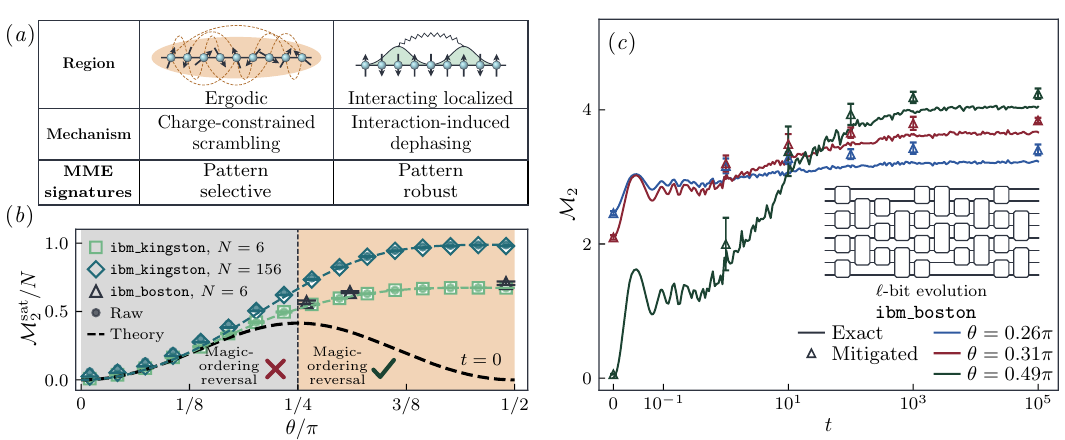}
  \caption{\textit{Interaction-induced dephasing and magic-ordering reversal.}
(a) Comparison of the mechanisms and initial-pattern dependence of the MME in the ergodic and interacting localized regimes. The localized regime is governed by interaction-induced dephasing and exhibits pattern-robust reversal, whereas the ergodic regime is pattern selective.
(b,c) Directly encoded interacting $\ell$-bit model [Eq.~\eqref{eq:Hlbits}], retaining terms through three-body order.
(b) Saturated magic density versus tilt $\theta$. The black dashed curve is the initial magic density, colored dashed curves are finite-size random-phase predictions, and symbols are hardware estimates on \texttt{ibm\_boston} and \texttt{ibm\_kingston}. For $\theta>\pi/4$, the initial and saturated magic orderings are reversed.
(c) Six-qubit magic dynamics for tilted ferromagnetic states with $\theta/\pi=0.26,0.31,0.49$. Solid curves are exact results and open triangles are mitigated Bell estimates on \texttt{ibm\_boston}. States with less initial magic attain more late-time magic. Inset: $\ell$-bit evolution circuit.}
  \label{fig:main_overview}
\end{figure*}

In this Letter, we establish interaction-induced dephasing as a mechanism for the MME in localized many-body systems [Fig.~\ref{fig:main_overview}(a)]. Using the effective $\ell$-bit description of many-body localization~\cite{Serbyn2013LocalConservation,Huse2014Phenomenology,Serbyn2013UniversalDephasing,Serbyn2014QuantumQuenches}, a random-phase analysis predicts that states with less initial magic attain higher saturated magic, with the same magic-ordering reversal for tilted ferromagnetic, N\'eel, and domain-wall states. Exact simulations further show that the states studied saturate earlier. We observe the predicted reversal on IBM superconducting processors~\cite{Kandala2017HardwareEfficient,Kandala2019ErrorMitigation,Koh2022ChiralDynamics,Kim2023QuantumUtility}: an exactly implemented $\ell$-bit evolution accesses long dephasing times at fixed circuit depth, while a microscopic spin model exhibits the reversal across all three initial-state families in the localized regime but only selectively in the ergodic regime. These results establish a dephasing-driven MME that operates without transport or conventional thermalization.

\maintextheading{Interaction-induced dephasing and magic-ordering reversal} We illustrate the MME mechanism with the phenomenological $\ell$-bit model of MBL~\cite{Huse2014Phenomenology,Serbyn2013LocalConservation},
% Keep the preceding paragraph line spacing while resizing only the equation.
\begingroup
\edef\dephasingparagraphbaselineskip{\the\baselineskip}
\small
\setlength{\baselineskip}{\dephasingparagraphbaselineskip}
\begin{equation}
H_{\ell\text{-bit}}=\sum_i h_i\tau_i^z+\sum_{i<j}J_{ij}\tau_i^z\tau_j^z+\sum_{i<j<k}K_{ijk}\tau_i^z\tau_j^z\tau_k^z+\cdots ,
\label{eq:Hlbits}
\end{equation}
\endgroup
whose couplings decay exponentially with range. Each $\tau_i^z$ is conserved, while interactions make its precession depend on the other $\ell$-bit configurations, producing dephasing over a broad range of timescales~\cite{Serbyn2014QuantumQuenches,xfp5-hhs4}. Thus configuration probabilities remain fixed while relative phases evolve coherently, providing a setting in which magic can relax without transport.

For a pure state, we quantify magic by the second stabilizer R\'enyi entropy (2-SRE)~\cite{Leone2022SRE,StabilizerEntropiesAndNonstabilizernessMonotones,Leone2024Monotones,Bittel2026Operational},
\begin{equation}
\mathcal M_2=-\log_2\mathcal I_4,\qquad
\mathcal I_4=2^{-N}\sum_{P\in\mathcal P_N}
|\langle\psi|P|\psi\rangle|^4 .
\label{eq:main_sre_definition}
\end{equation}
Identifying $\tau_i^z=Z_i$, we take
$\ket{\psi_0(\theta)}=\bigotimes_iR_y(\alpha_i)\ket{\uparrow}$,
with $\alpha_i=\theta$ or $\pi-\theta$ according to a binary pattern $s_i=0$ or $1$. The choices $0^N$, $(01)^{N/2}$, and $0^{N/2}1^{N/2}$ define tilted ferromagnetic (TFS), N\'eel (TNS)~\cite{LackOfSymmetryRestorationAfterAQuantum}, and domain-wall (TDWS) states. All three have the same initial magic density,
\begin{equation}
\frac{\mathcal M_2(0,\theta)}{N}
=-\log_2\!\left[1-\frac14\sin^2(2\theta)\right],
\label{eq:main_initial_magic_density}
\end{equation}
which decreases with $\theta$ for $\pi/4<\theta<\pi/2$ [Fig.~\ref{fig:main_overview}(b)]. By contrast, increasing $\theta$ makes the conserved configuration weights more uniform; at $\theta=\pi/2$, the zero-magic state $\ket{+}^{\otimes N}$ has uniform weights. Hence less initial magic need not imply less capacity for dephasing-generated magic.

Under Eq.~\eqref{eq:Hlbits}, only the configuration phases evolve. Replacing their late-time values by independent uniform phases while retaining the conserved weights gives, for fixed nonzero tilt and large $N$,
$\overline{\mathcal I_4}\simeq4[(1+\cos^4\theta)/2]^N$ and therefore
\begin{equation}
\lim_{N\to\infty}\frac{\mathcal M_2^{\mathrm{sat}}(\theta)}{N}
=1-\log_2(1+\cos^4\theta).
\label{eq:main_saturated_density}
\end{equation}
The finite-size expression used in Fig.~\ref{fig:main_overview}(b) is given in SI Sec.~\ref{sec:supp_lbit_theory}. The random-phase result is exact for this phase ensemble and describes Hamiltonian dynamics when the dephased moments self-average.

Crucially, for $\pi/4<\theta<\pi/2$, the initial magic density decreases with $\theta$ whereas the saturated density increases. Thus a state with less initial magic acquires more magic at saturation, with the endpoint ordering independent of the TFS, TNS, or TDWS pattern [Fig.~\ref{fig:main_overview}(b)]. The dynamical Mpemba criterion additionally requires this initially less resourceful state to reach its own higher saturated value sooner. Exact $N=14$ simulations confirm
$t_{\mathrm{ss}}(0.50\pi)<t_{\mathrm{ss}}(0.35\pi)$ for all studied initial-state families in the interacting $\ell$-bit model and random-field, quasiperiodic, and Stark localized spin chains [End Matter Fig.~\ref{fig:endmatter_localized_steady_state_time}; SI Sec.~\ref{sec:supp_numerics}].

Interactions are essential to this mechanism. In the Anderson limit $J_{ij}=K_{ijk}=0$, the dynamics factorizes into single-spin precession and randomizes only $N$ local phases rather than $2^N$ configuration phases. For the states and parameters studied, no magic-ordering reversal occurs, in contrast to the interacting case (End Matter and SI Sec.~\ref{sec:supp_lbit_theory}).

\begin{figure}[!b]
  \centering
  \includegraphics[width=0.98\columnwidth]{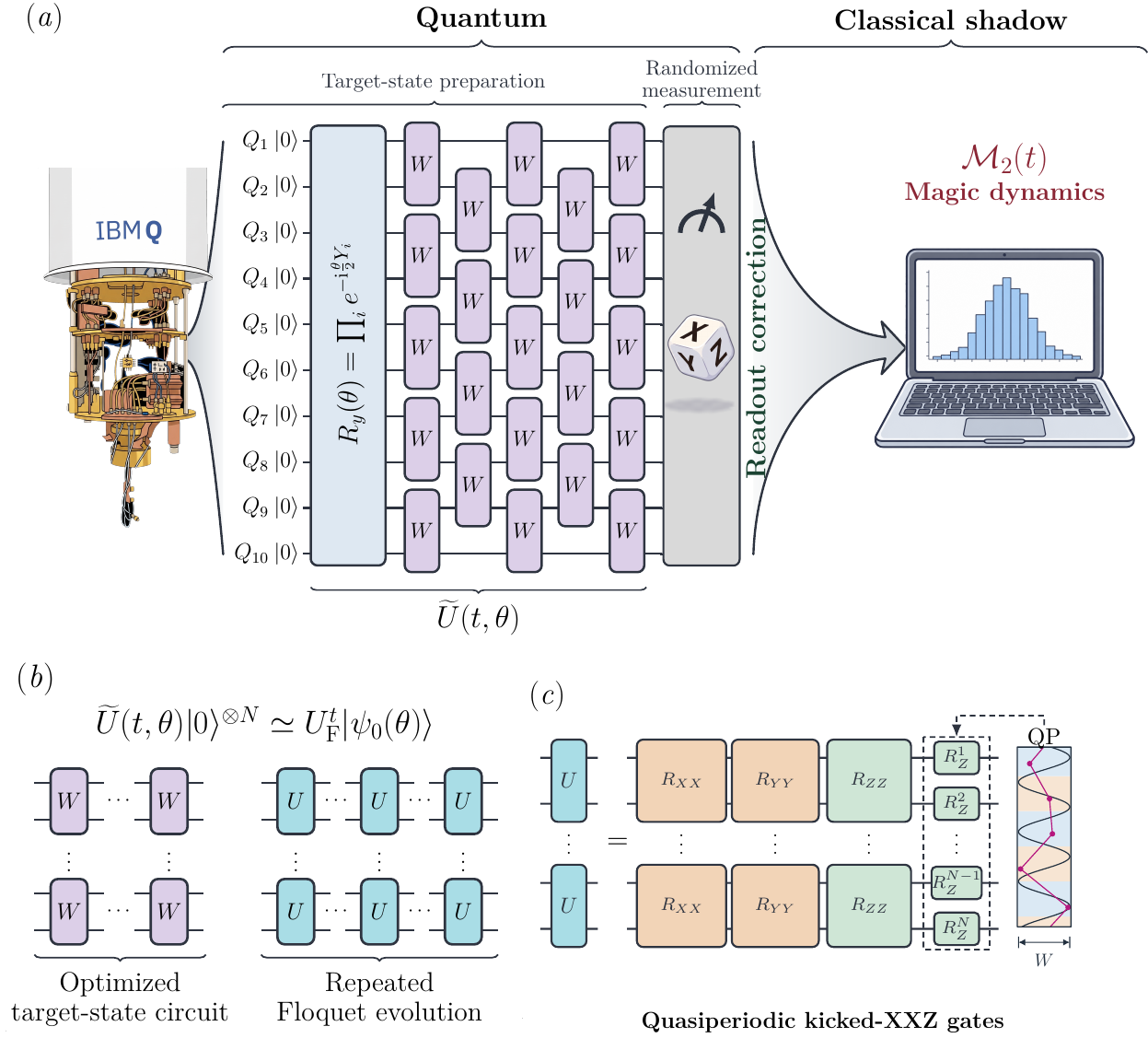}
  \caption{\textit{Short-circuit realization of the kicked XXZ chain} [Eq.~\eqref{eq:main_kicked_hamiltonian}] on \texttt{ibm\_boston}.
(a) Tilted ferromagnetic, N\'eel, or domain-wall states are prepared, evolved, and measured using randomized local Pauli bases to obtain $\mathcal M_2(t)$ with readout correction.
(b) For each sampled time, an optimized circuit $\widetilde U(t,\theta)$ prepares an approximation to $U_{\mathrm F}^{t}\ket{\psi_0(\theta)}$ from $\ket{0}^{\otimes N}$, avoiding repeated Floquet evolution.
(c) Elementary exchange and longitudinal gates of the quasiperiodically kicked XXZ drive. $W$ denotes optimized two-qubit blocks and QP the quasiperiodic field.}
  \label{fig:hardware_workflow}
\end{figure}

\begin{figure*}[!t]
  \centering
  \includegraphics[width=0.86\textwidth]{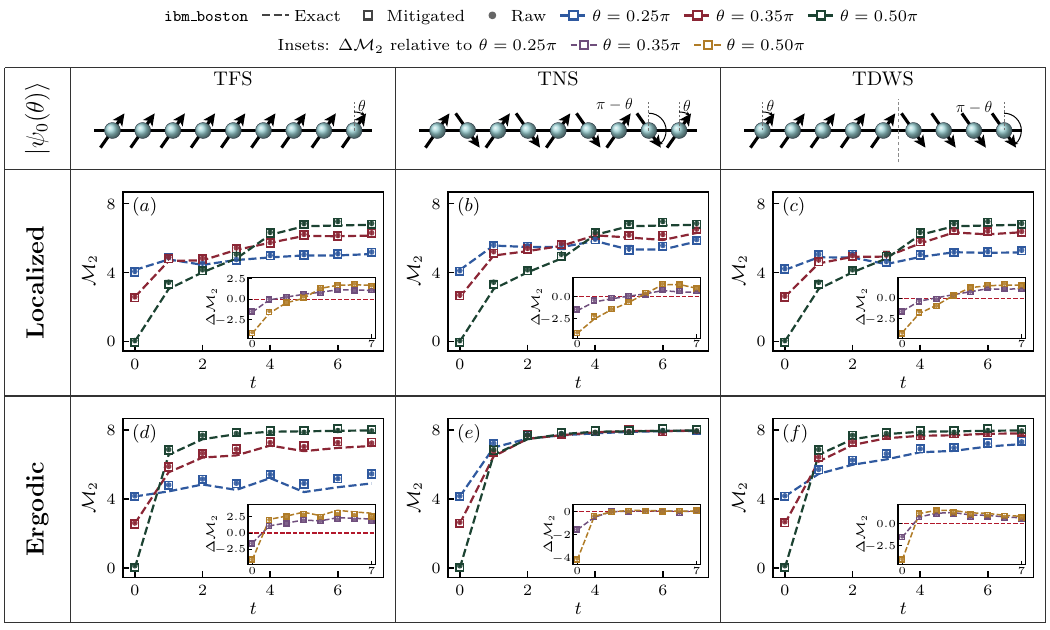}
  \caption{\textit{Magic Mpemba signatures in localized microscopic dynamics.}
Ten-qubit kicked-XXZ dynamics [Eq.~\eqref{eq:main_kicked_hamiltonian}]
on \texttt{ibm\_boston}.
All three studied initial-state families exhibit magic-ordering reversals in the
localized regime, whereas the ergodic control reverses only for TFS and TDWS
under the same $U(1)$ symmetry, supporting the dephasing picture beyond the
effective $\ell$-bit model.
$\mathcal M_2$ versus the number of Floquet periods $t$:
($a$--$c$) localized, $(K_1,K_2)=(0.201,0.250)$;
($d$--$f$) ergodic, $(K_1,K_2)=(0.511,0.637)$.
Remaining parameters and spectral diagnostics are given in
SI Sec.~\ref{sec:supp_hardware_execution} and
Table~\ref{tab:supp_kicked_parameters}.
Columns show TFS, TNS, and TDWS; colors denote
$\theta/\pi=0.25,0.35,0.50$.
Dashed curves are noiseless circuit predictions; filled circles and open
squares are raw and readout-mitigated estimates, with 95\%
hierarchical-bootstrap confidence intervals.
Insets: $\Delta\mathcal M_2(t,\theta)=\mathcal M_2(t,\theta)-\mathcal M_2(t,0.25\pi)$
for $\theta=0.35\pi$ (purple) and $0.50\pi$ (gold).
Crossings of the red dashed zero line from negative to positive mark
magic-ordering reversals.}
  \label{fig:main_kicked_l10}
\end{figure*}

\maintextheading{Quantum-hardware observation in the $\ell$-bit model} On quantum hardware, resolving slow dephasing requires
accurate measurements of a global fourth-order Pauli moment
over long simulated times. We encode time directly in the
gate angles of the diagonal $\ell$-bit propagator, implementing
the evolution without Trotterization and with a circuit depth
independent of simulated time. The number of two- and
three-body phase operations nevertheless grows with system size.
We implement the interacting model on \texttt{ibm\_boston},
using optimized phase-gate synthesis to reduce redundant
entangling operations~\cite{Amy2019CNOTPhase}
(SI Sec.~\ref{sec:supp_hardware_execution}).

Our dynamical measurements use two identically evolved
six-qubit registers. Pairwise Bell measurements recover
\begin{equation}
\mathcal I_4(t)=2^N\sum_{\bm r}p_{\bm r}(t)^2
\end{equation}
from the collision statistics of the joint outcomes
$p_{\bm r}$~\cite{ScalableMeasuresOfMagicResourceForQuantum}.
For ideal pure copies, the collision probability scales as
$2^{-N-\mathcal M_2(t)}$, making the measurement progressively
more demanding as either system size or magic increases.
Errors in the evolved copies and Bell measurement can also
produce an apparent increase in magic. We therefore combine
readout correction with zero-noise extrapolation (ZNE), which
reduces this bias and allows the predicted magic-ordering reversal
to be resolved during long-time dephasing
[Fig.~\ref{fig:main_overview}(c);
SI Sec.~\ref{sec:supp_hardware_execution}].

For $\theta/\pi=0.26$, $0.31$, and $0.49$, the measured
trajectories show that states with less initial magic acquire
more late-time magic [Fig.~\ref{fig:main_overview}(c)],
consistent with the ordering predicted by the random-phase
analysis. Error mitigation reduces the noise-induced upward
bias without removing the reversal. The late-time Bell
measurements are also consistent with the finite-size
random-phase prediction for saturated magic
[Fig.~\ref{fig:main_overview}(b)].

As a control, we implement the noninteracting Anderson limit
on the same processor using the same initial states and field
distribution, but with $J_{ij}=K_{ijk}=0$. No magic-ordering
reversal is observed over the measured time window
(see End Matter). The reversal therefore survives error
mitigation but disappears when interactions are removed,
providing a hardware-level control consistent with the
interaction-induced dephasing mechanism identified by our
random-phase analysis.

Fig.~\ref{fig:main_overview}(b) further includes a
model-assisted reference at $N=156$. Initial polarizations
$m_i=\langle Z_i\rangle$ measured for tilted product states on
\texttt{ibm\_kingston} are inserted into the finite-size
random-phase expression for $\mathcal M_2^{\mathrm{sat}}$.
This reference does not involve interacting evolution or a
direct measurement of saturated magic; instead, it tests the
predicted tilt dependence of the random-phase expression at
the full processor size
(SI Sec.~\ref{sec:supp_hardware_execution}).

\maintextheading{Microscopic realization and ergodic contrast} The $\ell$-bit evolution above isolates the dephasing mechanism.
We next ask whether the magic-ordering reversal survives noncommuting
microscopic interactions and how it differs from its counterpart
in chaotic dynamics. To this end, we implement a ten-qubit
quasiperiodically kicked XXZ chain whose localized regime is
expected to admit an effective interacting $\ell$-bit description,
despite the noncommuting physical-spin interactions. Localized
and ergodic dynamics are selected solely by the gate angles within
the same $U(1)$-symmetric nearest-neighbor circuit architecture.
Unlike the ensemble-averaged random circuits of
Ref.~\cite{Xiao2026NonstabilizernessMpemba}, the deterministic
drive avoids averaging over random circuit realizations and thereby
reduces the hardware overhead.

With unit drive period, the pulsed Hamiltonian is
\begin{equation}
\begin{aligned}
H_{\mathrm{KXXZ}}(t)
&=\sum_{m,a}\Bigl[\sum_{\mu=\mathrm e,\mathrm o}
\delta(t-m-\tau_{a,\mu})H_{\mu,a}\\
&\qquad+\delta(t-m-\tau_{a,\mathrm k})H_{\mathrm{kick},a}\Bigr],\\
H_{\mu,a}
&=\frac{K_a}{2}\sum_{(i,i+1)\in\mathcal B_\mu}
(X_iX_{i+1}+Y_iY_{i+1}),\\
H_{\mathrm{kick},a}
&=J_{z,a}\sum_{i=0}^{N-2}Z_iZ_{i+1}
+\sum_{i=0}^{N-1}h_i^{(a)}Z_i .
\end{aligned}
\label{eq:main_kicked_hamiltonian}
\end{equation}
Here $m\in\mathbb Z$ labels Floquet periods, $a=1,2$ labels
substeps, and $\mathcal B_{\mathrm e,o}$ denote the even- and
odd-bond sets of an open chain. The quasiperiodic fields are
$h_i^{(a)}=h_{0,a}+H_{\mathrm{qp},a}
\cos(2\pi\beta i+\varphi)$, with
$\beta=(\sqrt5-1)/2$. Each substep applies even-bond exchange,
odd-bond exchange, and a longitudinal kick
[Fig.~\ref{fig:hardware_workflow}(c)], giving
$U_a=e^{-\ii H_{\mathrm{kick},a}}
e^{-\ii H_{\mathrm o,a}}e^{-\ii H_{\mathrm e,a}}$
and $U_{\mathrm F}=U_2U_1$. Every pulse conserves the total
magnetization exactly. The localized and ergodic regimes,
diagnosed independently from the level statistics of $U_{\mathrm F}$,
therefore differ only through the gate angles, enabling a controlled
comparison of their magic-generation dynamics
(SI Sec.~\ref{sec:supp_hardware_execution}).

Directly repeating Floquet periods would accumulate gate errors and
could obscure the magic-ordering reversal. We instead prepare the
time-evolved states at the sampled times using independently
optimized state-preparation circuits
~\cite{Khatri2019QuantumCompiling,Koh2022ChiralDynamics,
Shen2025SPT,Chen2023AKLT}
[Fig.~\ref{fig:hardware_workflow}(b)]. Their fidelities and
magic errors are benchmarked in
SI Sec.~\ref{sec:supp_hardware_execution}.
We estimate the global fourth-order Pauli moment using randomized
local Pauli measurements without a second coherent copy
~\cite{Oliviero2022MeasuringMagic,Huang2020ClassicalShadows,
Elben2023RandomizedMeasurementToolbox,Joshi2025FullCountingStatistics}.
A Walsh--Hadamard evaluation avoids explicit enumeration of shot
quadruples, while readout correction is applied before the nonlinear
estimator. Purity normalization,
$\mathcal M_2=-\log_2[\Ifour(\rho)/\operatorname{Tr}(\rho^2)]$,
reduces mixedness bias, and a paired hierarchical bootstrap
quantifies the uncertainty in differences between trajectories
(SI Sec.~\ref{sec:supp_hardware_execution}).
For $N=10$, we use $(K_1,K_2)=(0.201,0.250)$ for the localized
data and $(0.511,0.637)$ for the ergodic control; the remaining
parameters and Floquet spectral diagnostics are given in
SI Sec.~\ref{sec:supp_hardware_execution} and
Table~\ref{tab:supp_kicked_parameters}.
In the localized regime, states with less initial magic overtake
those with more for TFS, TNS, and TDWS alike, as indicated by the
sign changes of $\Delta\mathcal M_2$
[Fig.~\ref{fig:main_kicked_l10}(a)--(c)].
At fixed tilt, the three families exhibit nearly coincident
trajectories, demonstrating a pattern-robust response across the
studied initial states, consistent with the $\ell$-bit prediction
despite the noncommuting microscopic interactions.

% Keep the concluding sentence from leaving a lone line on the following page.
\begingroup\widowpenalty=10000
The ergodic control behaves qualitatively differently. Although it
obeys the same $U(1)$ symmetry, magic-ordering reversals occur only for
TFS and TDWS, whereas TNS approaches a common late-time value
without a detectable reversal
[Fig.~\ref{fig:main_kicked_l10}(d)--(f)], consistent with
charge-conserving chaotic dynamics
~\cite{Xiao2026NonstabilizernessMpemba}.
Because TNS and TDWS have the same total-charge distribution,
their different behavior in the ergodic regime, also reproduced
at $N=6$ (SI Fig.~\ref{fig:supp_kicked_l6}), shows that the
conserved charge alone does not determine which initial-state
families exhibit the reversal. For the tilted states studied,
the effective $\ell$-bit description instead relates the reversed
saturated ordering to the conserved configuration weights and
interaction-induced dephasing, information not fixed by the
\mbox{total-charge distribution}.
\par\endgroup

% Keep the opening lines together at the page transition.
\begingroup\clubpenalty=10000
Taken together, the effective and microscopic implementations
constitute, to our knowledge, the first quantum-hardware
realization of MME signatures in localized many-body dynamics.
They extend the phenomenon beyond chaotic dynamics into a regime
where transport is suppressed. The reversal across all three
studied initial-state families in the localized regime, contrasted
with the pattern-selective ergodic control under the same
$U(1)$ symmetry, supports interaction-induced dephasing as a
distinct mechanism for the observed magic-ordering reversal.
\par\endgroup

The crossing itself establishes that a state with less initial
magic can acquire more magic at the same evolution time. The full
MME criterion further requires that this initially less resourceful
state reach its own higher saturated value sooner, despite starting
farther from it. Exact simulations establish this saturation-time
inversion for the studied states in the interacting $\ell$-bit
model and in localized static spin chains
[End Matter Fig.~\ref{fig:endmatter_localized_steady_state_time};
SI Sec.~\ref{sec:supp_numerics}].

\maintextheading{Discussion and outlook} Our results identify interaction-induced dephasing as a distinct
mechanism for the magic Mpemba effect in localized many-body
systems, demonstrating that anomalously accelerated magic
generation requires neither transport nor conventional
thermalization. The agreement between the random-phase analysis,
quantum-hardware magic-ordering reversals, and numerical saturation-time
inversion supports this picture in both the commuting $\ell$-bit
model and noncommuting microscopic spin dynamics.

More broadly, ``cooling-side'' Mpemba effects admit a unified
resource-theoretic description
~\cite{ResourceTheoreticalUnificationOfMpembaEffects,
aditya2025mpembaeffectsquantumcomplexity}, whereas no analogous
framework is yet established for anomalous resource generation.
Whether ``heating-side'' Mpemba effects across different quantum
resources can be understood within a common resource-theoretic
structure is therefore a natural open question. Systems with
constrained thermalization, including quantum many-body scars and
Hilbert-space fragmentation
~\cite{Turner2018Scars,Serbyn2021WeakBreaking,
Shen2024FockSkinScars,Sala2020Fragmentation,Khemani2020Shattering,
Moudgalya2022Review,PhysRevLett.130.120403,
regnault2025integercharacteristicpolynomialfactorization,
hao2026interacting,hao2026hilbert,PhysRevB.110.165109},
provide further settings in which dynamical constraints and
restricted exploration of Hilbert space may qualitatively reshape
resource generation. Extending such anomalously accelerated
generation beyond magic and entanglement
~\cite{bao2026entanglementmpembaeffect,
benjadi2026exponentialspeedupentanglementgeneration,
Zhang2026EntanglementEmbezzlement}
could reveal broader principles for quantum-resource dynamics and
resource-state preparation.

On quantum hardware, the present measurements resolve
the magic-ordering reversal, while directly determining which state reaches
its own saturated value sooner remains an important next step. The
fixed-depth $\ell$-bit implementation is naturally suited to such
long-time measurements. Mixed-state magic criteria and purity bounds
~\cite{Liu2026TriangleCriterion,Liu2026MagicGeometry}
further provide a route toward certifying magic directly in noisy
hardware states.

% Restore the class spacing for the remaining material.
\setlength{\abovecaptionskip}{\mainSavedAboveCaption}
\setlength{\belowcaptionskip}{\mainSavedBelowCaption}
\setlength{\abovedisplayskip}{\savedAboveDisplay}
\setlength{\belowdisplayskip}{\savedBelowDisplay}
\setlength{\abovedisplayshortskip}{\savedAboveShortDisplay}
\setlength{\belowdisplayshortskip}{\savedBelowShortDisplay}
\setlength{\jot}{\savedJot}
\setlength{\textfloatsep}{\savedTextFloatSep}
\setlength{\dbltextfloatsep}{\savedDblTextFloatSep}
\setlength{\intextsep}{\savedInTextSep}
\setlength{\floatsep}{\savedFloatSep}
\setlength{\dblfloatsep}{\savedDblFloatSep}

\begingroup\clubpenalty=10000
\maintextheading{Acknowledgments}
C.~H.\ Lee acknowledges support from the Ministry of Education, Singapore
(MOE Tier-II award numbers MOE-T2EP50224-0007 and MOE-T2EP50125-0019). S.\ Liu was supported by the Gordon and Betty Moore Foundation (grant GBMF8685) towards the Princeton theory program, the Gordon and Betty Moore Foundation’s EPiQS Initiative (grant GBMF11070), the Global Collaborative Network Grant at Princeton University, the Simons Investigator Grant No. 404513, the NSF MRSEC (grant DMR-2011750), the Simons Collaboration on New Frontiers in Superconductivity (grant SFI-MPS-NFS-00006741-01), the Princeton Catalysis Initiative (PCI), the Schmidt Transformative Technology Fund at Princeton University. J.-X.\ Zhong was supported by the National Natural Science Foundation of China
(Grant Nos.~12374046 and 11874316), the Shanghai Science and Technology
Innovation Action Plan (Grant No.~24LZ1400800), the National Basic Research
Program of China (Grant No.~2015CB921103), and the Program for Changjiang
Scholars and Innovative Research Teams in Universities (Grant No.~IRT13093).
H.-Z.\ Li is supported by the China Scholarship Council Scholarship (Grant
No.~202506890103).
The authors acknowledge the use of IBM Quantum services for this work.
The views expressed are those of the authors, and do not reflect the official
policy or position of IBM or the IBM Quantum team.
\par\endgroup

\makeatletter
\let\savedaddcontentsline\addcontentsline
\let\addcontentsline\@gobblethree
% Avoid an oversized gap before the author notes and references.
\let\savedmainbibsection\bibsection
\patchcmd{\bibsection}{\baselineskip26\p@}{\baselineskip14\p@}{}{}
\patchcmd{\bibsection}{\addvspace{19\p@}}{\addvspace{8\p@}}{}{}
% Keep the author line at the final balanced column break with its continuation.
\let\savedmainlbibitem\@lbibitem
\def\@lbibitem[#1]#2{%
  \savedmainlbibitem[#1]{#2}%
  \ifstrequal{#2}{Shen2024FockSkinScars}
    {\clubpenalty=10000\@clubpenalty=10000\relax}%
    {\clubpenalty=5000\@clubpenalty=5000\relax}%
  % Flexible entry gaps align the two column bottoms on the final reference page.
  \ifnum\c@page=9\relax\itemsep=0pt plus 3pt\relax\fi
}
\microtypesetup{expansion=false}
\bibliography{ref}
\let\@lbibitem\savedmainlbibitem
\let\bibsection\savedmainbibsection
\microtypesetup{expansion=true}
\let\addcontentsline\savedaddcontentsline
\makeatother

% Balance the last bibliography page before starting End Matter.
\onecolumngrid
\clearpage
\renewcommand{\topfraction}{0.92}
\renewcommand{\bottomfraction}{0.6}
\renewcommand{\textfraction}{0.06}
\renewcommand{\floatpagefraction}{0.8}

\begingroup
\compactfloatspacing
\compactmathspacing

\noindent\makebox[\textwidth][c]{\textbf{\large End Matter}}\par
\vspace{5pt}

\begin{figure}[H]
    \centering
    \includegraphics[width=0.90\textwidth]
    {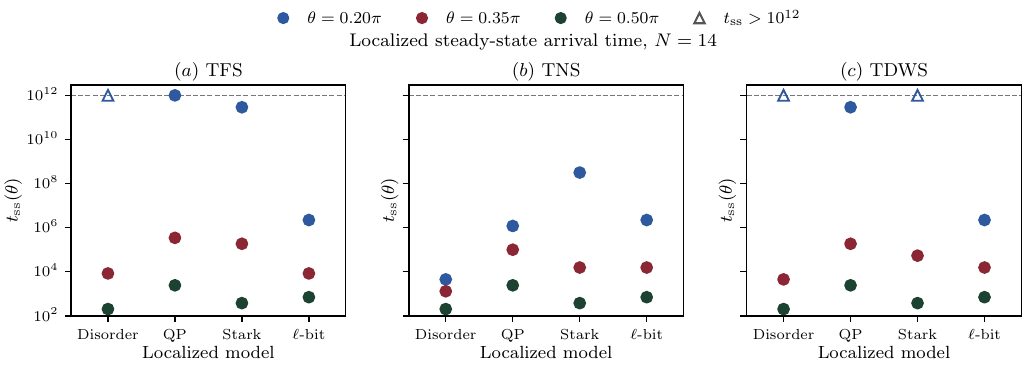}
    \caption{\textit{Numerical saturation-time inversion across localized models.}
    Saturation times $t_{\mathrm{ss}}$ at $N=14$ for
    (a) TFS, (b) TNS, and (c) TDWS.
    Colors denote $\theta/\pi=0.20,0.35,0.50$.
    The localized models are random-disorder and quasiperiodic XXZ
    ($W=10$), Stark XXZ ($F=6$), and the interacting three-body
    $\ell$-bit model ($\xi=0.5$).
    The saturation criterion uses relative tolerance $\epsilon=0.1$.
    Open triangles indicate lower bounds
    $t_{\mathrm{ss}}>10^{12}$; the dashed line marks the observation
    limit $t_{\max}=10^{12}$.}
    \label{fig:endmatter_localized_steady_state_time}
\end{figure}

\twocolumngrid

\maintextheading{Saturation-time inversion across localized models} The magic-ordering reversals in Figs.~\ref{fig:main_overview}(c) and
\ref{fig:main_kicked_l10} establish that states with less initial
magic can overtake states with more, but a crossing alone does not
determine which state reaches its own saturated value sooner.
The full dynamical MME criterion additionally requires the initially
less resourceful state to reach its own asymptotic value earlier,
despite beginning farther from it. We test this criterion using exact
$N=14$ dynamics of the interacting $\ell$-bit model
[Eq.~\eqref{eq:Hlbits}] and static XXZ chains with random,
quasiperiodic, and Stark fields. For the microscopic XXZ models,
magic is evaluated in the physical-spin basis. We consider the same
tilted ferromagnetic, N\'eel, and domain-wall initial-state families
used throughout the main text.

Because the saturated magic depends on the initial state, different
trajectories should be compared relative to their own asymptotic
values. We therefore use the normalized distance
$\Delta_\theta(t)=
|\mathcal M_2(t,\theta)-\mathcal M_2^{\mathrm{sat}}(\theta)|
/
|\mathcal M_2(0,\theta)-\mathcal M_2^{\mathrm{sat}}(\theta)|$,
for which $\Delta_\theta(0)=1$.
The saturation time $t_{\mathrm{ss}}$ is defined as the earliest
sampled time from which $\Delta_\theta(t)<0.1$ at every remaining
sample up to $t_{\max}=10^{12}$. Requiring the threshold to remain
satisfied excludes transient crossings caused by finite-size
oscillations. The saturated value is estimated separately for each
trajectory from its late-time window; varying this window does not
change the time ordering reported below
(SI Sec.~\ref{sec:supp_numerics}).

Fig.~\ref{fig:endmatter_localized_steady_state_time}(a)--(c)
shows $t_{\mathrm{ss}}(0.50\pi)<t_{\mathrm{ss}}(0.35\pi)$
for every localized model and initial-state family studied.
This inversion is stronger than the magic-ordering reversal alone:
the $\theta=0.50\pi$ state starts with less magic and farther from
its own saturated value, yet reaches a higher saturated magic
earlier. Its faster arrival therefore cannot be attributed simply
to a shorter distance from the final state. At $\theta=\pi/2$,
all three initial-state families reduce to the common zero-magic
state $\ket{+}^{\otimes N}$; the three panels compare its relaxation
with those of the distinct lower-tilt TFS, TNS, and TDWS states.

The same saturation-time inversion occurs in four qualitatively
different localized settings: the effective interacting $\ell$-bit
model, random-disorder localization, quasiperiodic localization,
and Stark localization. The microscopic models contain
noncommuting physical-spin interactions, in contrast to the
commuting directly encoded $\ell$-bit Hamiltonian. The common
time ordering across these models and across all three studied
initial-state patterns therefore shows that the dynamical MME
signature is not tied to a particular disorder realization,
localization mechanism, or commuting effective description.
Hamiltonian parameters, localization diagnostics, numerical
procedures, and full trajectories are given in
SI Sec.~\ref{sec:supp_numerics}.

\maintextheading{Role of interactions in the magic-ordering reversal} The interacting $\ell$-bit model separates transport from many-body
dephasing: every $\tau_i^z$ is conserved, whereas the interaction
terms generate configuration-dependent phases. To isolate the role
of these interactions, we set $J_{ij}=K_{ijk}=0$ in the implemented
Hamiltonian [Eq.~\eqref{eq:Hlbits}], reducing the evolution to
independent spin precession under $H_0=\sum_i h_i Z_i$.
We retain the same field distribution, initial states, processor,
and Bell-measurement protocol as in
Fig.~\ref{fig:main_overview}(c), so that the comparison removes
the interaction terms while leaving the remaining hardware
setting unchanged.

In this Anderson limit, the state remains a product state throughout
the evolution. Local precession nevertheless generates nonzero
magic, so the absence of interactions does not imply an absence of
resource generation. The distinction instead lies in the phase
structure. Independent spin precession produces only local phases,
whereas interacting $\ell$-bit evolution generates
configuration-dependent many-body phases across the computational
basis. Correspondingly, the two limits produce qualitatively
different late-time magic. At $\theta=\pi/2$, for example, the
local-phase-averaged magic density is approximately $0.19$, whereas
the interacting random-phase result approaches one bit per qubit
in the thermodynamic limit
[Eq.~\eqref{eq:main_saturated_density};
SI Sec.~\ref{sec:supp_lbit_theory}].

Fig.~\ref{fig:endmatter_anderson} shows the measured
noninteracting trajectories. Magic grows under local precession,
but the ordering of the three initial states does not reverse over
the measured time window. Magic generation by itself is therefore
insufficient to account for the observed Mpemba magic-ordering reversal.
By contrast, the otherwise matched interacting implementation in
Fig.~\ref{fig:main_overview}(c) exhibits the predicted reversal.
Together, these measurements support the interpretation that the
observed effect originates from interaction-induced many-body
dephasing rather than from localization or local precession alone.
Readout correction is applied without ZNE; implementation details
and analytical expressions for the Anderson limit are given in
SI Secs.~\ref{sec:supp_hardware_execution} and
\ref{sec:supp_lbit_theory}, respectively.

\newpage
\begin{figure}[H]
    \centering
    \includegraphics[width=\columnwidth]
    {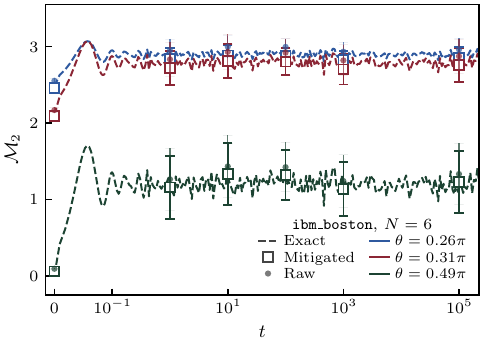}
    \caption{\textit{Magic dynamics without interactions on}
    \texttt{ibm\_boston}.
    Six-qubit Bell estimates in the Anderson limit,
    $J_{ij}=K_{ijk}=0$, using the same initial states and field
    distribution as Fig.~\ref{fig:main_overview}(c).
    Dashed curves are exact trajectories, filled circles are raw
    estimates, and open squares are readout-corrected estimates.
    Magic is generated, but no magic-ordering reversal occurs over the
    measured window.
    Error bars denote one standard deviation over 15 field
    realizations. The first tick denotes $t=0$; the remaining
    time axis is logarithmic.}
    \label{fig:endmatter_anderson}
\end{figure}

\par
\onecolumngrid
\endgroup

\renewcommand{\topfraction}{0.7}
\renewcommand{\bottomfraction}{0.3}
\renewcommand{\textfraction}{0.2}
\renewcommand{\floatpagefraction}{0.5}
\clearpage
\onecolumngrid
\begingroup
\makeatletter
\let\set@footnotewidth\set@footnotewidth@one
\let\compose@footnotes\compose@footnotes@one
\renewcommand{\citenumfont}[1]{S#1}
\renewcommand{\bibnumfmt}[1]{[S#1]}
\renewcommand{\@biblabel}[1]{[S#1]}
\makeatother

\begin{center}
  \textbf{\large Supplemental Information for\\``Observing the magic Mpemba effect in localized dynamics\\on a digital quantum computer''}
\end{center}

\setcounter{equation}{0}
\setcounter{figure}{0}
\setcounter{table}{0}
\renewcommand{\theequation}{S\arabic{equation}}
\renewcommand{\thefigure}{S\arabic{figure}}
\renewcommand{\thetable}{S\arabic{table}}
\renewcommand{\theHequation}{S\arabic{equation}}
\renewcommand{\theHfigure}{S\arabic{figure}}
\renewcommand{\theHtable}{S\arabic{table}}

\begin{center}
Han-Ze Li,$^{1,2,*}$ Xianquan Yan,$^{1,3,*}$ Yi-Rui Zhang,$^{2}$ Jian-Xin Zhong,$^{2,\dagger}$ Shuo Liu,$^{4,\ddagger}$ and Ching Hua Lee$^{1,\S}$
\vspace{6pt}

\textit{$^{1}$Department of Physics, National University of Singapore, Singapore 117551}\\
\textit{$^{2}$Institute for Quantum Science and Technology, Shanghai University, Shanghai 200444, China}\\
\textit{$^{3}$Department of Computer Science, National University of Singapore, Singapore 117417}\\
\textit{$^{4}$Department of Physics, Princeton University, Princeton, New Jersey 08544, USA}
\end{center}
\vspace{8pt}

\setcounter{section}{0}
\setcounter{subsection}{0}
\setcounter{subsubsection}{0}
\setcounter{secnumdepth}{2}
\setcounter{tocdepth}{2}
\renewcommand{\thesection}{S\arabic{section}}
\renewcommand{\thesubsection}{\Alph{subsection}}
\renewcommand{\thesubsubsection}{\arabic{subsubsection}}
\makeatletter
\renewcommand{\@seccntformat}[1]{\csname the#1\endcsname.\quad}
\makeatother
\renewcommand{\theHsection}{supp.\arabic{section}}
\renewcommand{\theHsubsection}{supp.\arabic{section}.\Alph{subsection}}
\renewcommand{\theHsubsubsection}{supp.\arabic{section}.\Alph{subsection}.\arabic{subsubsection}}
% Compact SI heading hierarchy, preserving the REVTeX heading styles.
\makeatletter
\renewcommand{\section}{%
  \@startsection{section}{1}{\z@}{12pt plus 2pt minus 1pt}{7pt}%
  {\normalfont\small\bfseries\centering}}
\renewcommand{\subsection}{%
  \@startsection{subsection}{2}{\z@}{10pt plus 2pt minus 1pt}{6pt}%
  {\normalfont\small\bfseries\centering}}
\renewcommand{\subsubsection}{%
  \@startsection{subsubsection}{3}{\z@}{8pt plus 1pt minus 1pt}{4pt}%
  {\normalfont\small\itshape\centering}}
\makeatother
\let\supplementarysection\section
\makeatletter
\renewcommand{\section}{%
  \@ifstar{\supplementarysection*}{\clearpage\supplementarysection}}
\makeatother

\startsupplementaryfloatlists

\noindent This Supplemental Information covers:
\begin{itemize}\setlength{\itemsep}{0pt}
  \item Direct $\ell$-bit and kicked-XXZ implementations and measurement protocols on quantum hardware.
  \item Six-qubit kicked-XXZ magic dynamics. 
  \item Theory of initial magic, saturated values after dephasing, and the Anderson limit.
  \item Numerical methods, localization diagnostics, and relaxation-time comparisons.
\end{itemize}

\tableofcontents

\raggedbottom
\compactmathspacing
\setlength{\textfloatsep}{8pt plus 1pt minus 1pt}
\setlength{\floatsep}{6pt plus 1pt minus 1pt}
\setlength{\intextsep}{6pt plus 1pt minus 1pt}
\setlength{\abovecaptionskip}{4pt}
\setlength{\belowcaptionskip}{0pt}
\setcounter{secnumdepth}{3}
\clearpage
\section{Quantum-simulation implementations on quantum hardware}
\label{sec:supp_hardware_execution}

The six-qubit $\ell$-bit trajectories in Fig.~\ref{fig:main_overview}(c)
and the noninteracting control in Fig.~\ref{fig:endmatter_anderson}
are acquired on \texttt{ibm\_boston}. The diagonal propagator encodes time
in rotation angles, allowing slow dephasing to be resolved at fixed
circuit depth. For the noncommuting kicked XXZ drive, short circuits
prepare individual target states on the same processor. The resulting
$N=10$ trajectories [Fig.~\ref{fig:main_kicked_l10}] and $N=6$ comparison
[Fig.~\ref{fig:supp_kicked_l6}] distinguish localized and ergodic magic
ordering across three initial-state patterns.

Polarization measurements on \texttt{ibm\_kingston} supply the weights
for the model-assisted saturated-magic references at $N=6$ and 156 in
Fig.~\ref{fig:main_overview}(b). These references use the random-phase
formula, without interacting evolution or direct saturated-magic
measurement at either size.

\subsection{Realization of \texorpdfstring{$\ell$}{l}-bit dynamics}
\label{sec:supp_lbit_realization}

\subsubsection{Target Hamiltonian and initial-state preparation}
\label{sec:supp_lbit_hamiltonian_preparation}

For the dynamics in Fig.~\ref{fig:main_overview}(c), we encode the
effective local integrals of motion ($\ell$-bits) directly in the processor
qubits, taking $\tau_i^z=Z_i$. The implemented Hamiltonian
[Eq.~\eqref{eq:Hlbits}] retains interactions through three-body order,
\begin{equation}
H_{\ell\text{-bit}}=
\sum_i h_i Z_i+
\sum_{i<j}J_{ij}Z_iZ_j+
\sum_{i<j<k}K_{ijk}Z_iZ_jZ_k.
\label{eq:supp_lbit_hardware_hamiltonian}
\end{equation}
For each disorder realization, $h_i$ is uniform on $[-15,15]$, while
$J_{ij}$ and $K_{ijk}$ are independent zero-mean Gaussian variables
with standard deviations $e^{-(j-i)/\xi}$ and $e^{-(k-i)/\xi}$,
respectively, and $\xi=0.5$.  At $N=6$, this gives 6 one-body,
15 two-body, and 20 three-body phase terms.  The Anderson comparison
uses the same field distribution with $J_{ij}=K_{ijk}=0$.

The time-resolved measurements use tilted ferromagnetic states (TFS).
Starting from $\ket{0}^{\otimes N}$, parallel $R_y(\theta)$ rotations
prepare $\ket{\psi_0(\theta)}=\bigotimes_iR_y(\theta)\ket{0}$,
where $Z\ket{0}=\ket{0}$ and $R_y(\theta)=e^{-\ii\theta Y/2}$.
We identify $\ket{0}=\ket{\uparrow}$ and label sites $i=0,\ldots,N-1$,
as in the main text.
We prepare this state on two registers and apply the same disorder
realization and evolution time to both.  Varying $\theta$ changes the
initial configuration weights while the Hamiltonian ensemble is held
fixed.  This allows the measured magic ordering to be compared under
interacting and noninteracting evolution.

\subsubsection{Exact synthesis of the diagonal evolution}
\label{sec:supp_lbit_phase_synthesis}

All terms in Eq.~\eqref{eq:supp_lbit_hardware_hamiltonian} commute.
With $R_P(\alpha)=e^{-\ii\alpha P/2}$ for a Pauli string $P$, the
propagator therefore factorizes exactly into one-, two-, and three-body
phase rotations,
\begin{equation}
\begin{gathered}
U_\ell(t)=e^{-\ii tH_{\ell\text{-bit}}}
=\prod_i R_{Z_i}(2th_i)
\prod_{i<j}R_{Z_iZ_j}(2tJ_{ij})
\prod_{i<j<k}R_{Z_iZ_jZ_k}(2tK_{ijk}),\\[1.2ex]
\includegraphics[width=0.94\textwidth]{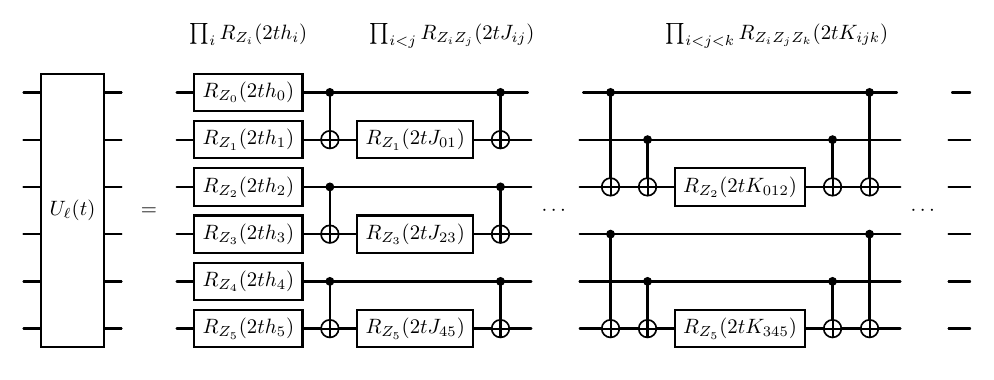}
\end{gathered}
\label{eq:supp_lbit_exact_evolution}
\end{equation}
The circuit shows representative disjoint pairs and triples; the ellipses
include all remaining $i<j$ and $i<j<k$ terms, not additional time steps.
In particular, the implemented Hamiltonian is not restricted to the
adjacent sites displayed in the schematic.

Each many-body rotation is implemented by computing and uncomputing a
binary parity.  On a computational-basis configuration $\ket{\bm x}$,
where $Z_i\ket{\bm x}=(-1)^{x_i}\ket{\bm x}$, controlled-NOT (CNOT)
gates encode $x_i\oplus x_j$ or $x_i\oplus x_j\oplus x_k$ on a target
qubit.  The target rotation $R_Z(\alpha)$ then contributes the phase
$\exp[-\ii\alpha(-1)^{x_i\oplus x_j}/2]$ or
$\exp[-\ii\alpha(-1)^{x_i\oplus x_j\oplus x_k}/2]$.
Reversing the CNOT sequence restores the original bit labels without
undoing this phase.  Writing
$C_{i\to j}=\mathrm{CNOT}_{i\to j}$, the two- and three-body identities are
\begin{equation}
\begin{aligned}
R_{Z_iZ_j}(\alpha)
&=C_{i\to j}R_{Z_j}(\alpha)C_{i\to j},\\
R_{Z_iZ_jZ_k}(\alpha)
&=C_{i\to k}C_{j\to k}R_{Z_k}(\alpha)
C_{j\to k}C_{i\to k}.
\end{aligned}
\label{eq:supp_lbit_parity_synthesis}
\end{equation}
Thus the many-body phases require only one-qubit rotations and two-qubit
entangling gates; the three-body term does not require a native
three-qubit gate.  Eq.~\eqref{eq:supp_lbit_exact_evolution} displays
these elementary identities rather than the optimized gate schedule.
We synthesize the complete phase polynomial with
GraySynth~\cite{SI-Amy2019CNOTPhase}, which reuses intermediate parities
instead of separately computing and undoing them for every term.  This
reduces redundant entangling operations while preserving the ideal
propagator.

Changing $t$ changes the rotation angles in
Eq.~\eqref{eq:supp_lbit_exact_evolution}; it does not require repeating
an evolution layer $t$ times.  This gives access to the sampled times up
to $t=10^5$ without a Trotter approximation or a target-state fidelity
optimization.  The exactness here concerns the gate decomposition of the
chosen three-body Hamiltonian.  Hardware gate errors and readout errors
remain and are treated below.  With $J_{ij}=K_{ijk}=0$, the same construction
reduces to independent $R_Z(2th_i)$ precession, giving the noninteracting
control in Fig.~\ref{fig:endmatter_anderson} for the interacting
trajectories in Fig.~\ref{fig:main_overview}(c).

\subsubsection{Two-copy Bell measurement and error mitigation}
\label{sec:supp_lbit_two_copy}
\label{sec:supp_lbit_bell_execution}

The magic estimates in Figs.~\ref{fig:main_overview}(c) and
\ref{fig:endmatter_anderson} use two-copy Bell measurements.
After preparing and evolving the copies, we apply
$\mathrm{CNOT}_{A_i\to B_i}$ to each pair of corresponding qubits,
followed by a Hadamard on $A_i$.  Computational-basis readout then
measures the Bell basis, as shown in
Fig.~\ref{fig:supp_lbit_bell_circuit}.  The upper and lower registers,
marked $a$ and $b$ in the figure, give the bit strings $\bm a$ and
$\bm b$ in each shot.  Both contain the same ideal state
$\ket{\psi(t,\theta)}=U_\ell(t)\ket{\psi_0(\theta)}$ before Bell
analysis.  No randomized measurement basis is needed in this scheme.

\begin{figure}[!htbp]
  \centering
  \includegraphics[width=0.50\textwidth]{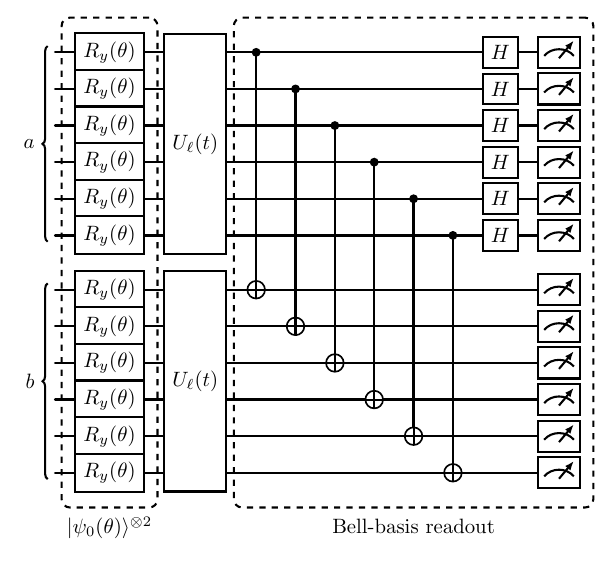}
  \caption[Two-copy Bell measurement of \(\mathcal M_2\).]
  {\textit{Two-copy Bell measurement of \(\mathcal M_2\).}
  Parallel $R_y(\theta)$ rotations prepare two TFS copies, followed by
  identical $U_\ell(t)$ evolution.  Pairwise CNOT gates and Hadamards
  on the upper register map Bell outcomes to the jointly measured
  bit strings $(\bm a,\bm b)$.}
  \label{fig:supp_lbit_bell_circuit}
\end{figure}

\begin{samepage}
For each qubit pair, the recorded bits $a_i$ and $b_i$ label the
eigenvalues $(-1)^{a_i}$ of $X_{A_i}X_{B_i}$ and $(-1)^{b_i}$ of
$Z_{A_i}Z_{B_i}$.  Equivalently, the measured basis and its probabilities
are
\begin{equation}
\begin{aligned}
\ket{\beta_{\bm a,\bm b}}
&=\bigotimes_{i=0}^{N-1}
\left(I_{A_i}\otimes X_{B_i}^{b_i}Z_{B_i}^{a_i}\right)
\frac{\ket{00}_{A_iB_i}+\ket{11}_{A_iB_i}}{\sqrt{2}},\\
p_{\bm a,\bm b}(t,\theta)
&=\left|\bra{\beta_{\bm a,\bm b}}
\ket{\psi(t,\theta)}^{\otimes2}\right|^2.
\end{aligned}
\label{eq:supp_bell_outcome_probability}
\end{equation}
\end{samepage}
Orthogonality of the Bell-basis parity signs relates the collision
probability to the fourth Pauli moment.  For the two identical pure
copies,
\begin{equation}
\mathcal I_4(t,\theta)
=2^N\sum_{\bm a,\bm b}p_{\bm a,\bm b}(t,\theta)^2
=2^{-N}\sum_{P\in\cP_N}
\left|\bra{\psi(t,\theta)}P\ket{\psi(t,\theta)}\right|^4,
\qquad
\mathcal M_2(t,\theta)=-\log_2\mathcal I_4(t,\theta).
\label{eq:supp_bell_collision_moment}
\end{equation}
Here $\cP_N=\{I,X,Y,Z\}^{\otimes N}$, and $\mathcal M_2$ is the
second stabilizer R\'enyi entropy (2-SRE)~\cite{SI-Leone2022SRE,SI-ScalableMeasuresOfMagicResourceForQuantum}.
With $S$ shots and counts $n_{\bm a,\bm b}$, the distinct-shot collision
estimate is $\sum_{\bm a,\bm b}n_{\bm a,\bm b}(n_{\bm a,\bm b}-1)/[S(S-1)]$.
This removes self-pairs from the empirical collision count; the subsequent
logarithm still has finite-shot bias.  For noisy outputs, a reduced Bell
collision probability produces an upward shift in this pure-state
diagnostic, not necessarily an increase in useful magic.\footnote{\parbox[t]{0.94\textwidth}{For the one-qubit stabilizer mixture
\(\rho_p=(1-p)\ket{+}\bra{+}+p\mathbb I/2\), the state remains a convex
mixture of stabilizer states, yet the naive two-copy formula returns
\(-\log_2\{[1+(1-p)^4]/2\}>0\) for every \(p>0\): 0.272 bits at \(p=0.1\)
and one bit for the maximally mixed state.}}
Readout errors and decoherence can broaden the Bell distribution and
generate this bias.  We therefore report raw and mitigated estimates
separately, with assignment correction and gate-noise extrapolation as
specified below.

The Bell data on \texttt{ibm\_boston} use 15 disorder realizations,
8000 shots per circuit, $\theta/\pi=0.26,0.31,0.49$, and
$t=0,1,10,10^2,10^3,10^5$.
Matrix-free measurement mitigation (M3)~\cite{SI-Nation2021M3}, with
4000 calibration shots, first corrects the joint readout distribution.
The resulting quasidistribution is projected onto its nearest probability
distribution before the collision is evaluated.  For interacting evolution
at $t>0$, selected native controlled-$Z$ ($\mathrm{CZ}$) gates are folded
as $\mathrm{CZ}\mapsto\mathrm{CZ}^3=\mathrm{CZ}$ to obtain
$\mathrm{CZ}$-count noise scales $\lambda_1=1$ and $\lambda_2=2$.
Linear zero-noise extrapolation (ZNE) then gives
$\mathcal M_2^{\mathrm{ZNE}}(t,\theta)
=[\lambda_2\mathcal M_2(t,\theta;\lambda_1)
-\lambda_1\mathcal M_2(t,\theta;\lambda_2)]/(\lambda_2-\lambda_1)$,
where the argument after the semicolon denotes the noise
scale~\cite{SI-Temme2017ZNE}.  Anderson data and interacting $t=0$ data
use M3 alone.  We evaluate the observable for each disorder realization
before averaging; the displayed error bars are one sample standard
deviation over the 15 realizations.

The full raw and mitigated trajectories underlying
Fig.~\ref{fig:main_overview}(c) are compared in
Fig.~\ref{fig:supp_lbit_raw_mitigated}.

\begin{figure}[!htbp]
  \centering
  \includegraphics[width=0.64\textwidth]{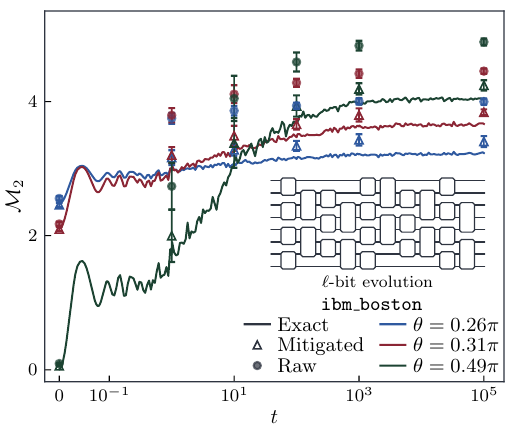}
  \caption[Raw and mitigated $\ell$-bit magic dynamics.]
  {\textit{Raw and mitigated $\ell$-bit magic dynamics.}
  Complete Bell-measurement comparison underlying main-text
  Fig.~\ref{fig:main_overview}(c): six-qubit TFS at
  $\theta/\pi=0.26,0.31,0.49$, with exact trajectories (solid curves),
  raw estimates (circles), and mitigated estimates (open triangles).
  Each circuit uses 8000 shots; error bars are one standard deviation over
  15 disorder realizations. Interacting $t>0$ data use M3 and ZNE;
  $t=0$ uses M3 alone. At late times, mitigation reduces the upward bias
  while preserving the magic-ordering reversal.}
  \label{fig:supp_lbit_raw_mitigated}
\end{figure}

\subsubsection{Polarization-based reconstruction of saturated magic}
\label{sec:supp_lbit_full_register}

Initial-state polarization measurements supply inputs to the
model-assisted references for saturated magic in the $\ell$-bit model
[Fig.~\ref{fig:main_overview}(b)]. Because the
diagonal evolution preserves configuration probabilities, the
phase-averaged prediction can be evaluated from the initial site
polarizations $m_i=\langle Z_i\rangle$.  For the product-weight model,
the finite-size expression is
$\overline{\mathcal I_4}
=4\prod_i[(1+m_i^4)/2]
-6\prod_i[(1+m_i^2)^2/4]
+3\prod_i[(1+6m_i^2+m_i^4)/8]$,
and the reconstructed saturated value is
$\mathcal M_2^{\mathrm{sat}}=-\log_2\overline{\mathcal I_4}$.
The axis label $\mathcal M_2^{\mathrm{sat}}$ in
Fig.~\ref{fig:main_overview}(b) denotes the measured or reconstructed
estimate of the saturated value.
For homogeneous ideal preparation, $m_i=\cos\theta$, this reduces to
Eq.~\eqref{eq:supp_random_phase_i4_exact}.  Using the individually
measured $m_i$ retains site-dependent preparation and readout variation.
The derivation and phase-averaging assumptions are given in
Sec.~\ref{sec:supp_lbit_theory}.

On \texttt{ibm\_kingston}, parallel $R_y(\theta)$ rotations followed
by $Z$ readout supply these polarizations for $N=6$ and 156.
The six-qubit subset is selected by its snapshot readout errors;
connectivity does not constrain this rotation-and-readout protocol.
We use 16 angles, $\theta/\pi=0.01,0.042,\ldots,0.49$, giving
32 data circuits across the two sizes.  Four all-up or all-down calibration
circuits in the same job determine the assignment correction.
Each data and calibration circuit uses 4000 shots.  The raw polarization
$r_i$ is corrected as
$m_i=[r_i+p_i(1|0)-p_i(0|1)]/[1-p_i(1|0)-p_i(0|1)]$,
clipped to $[-1,1]$, and inserted into the expression for the saturated value;
$p_i(b|a)$ is the probability of reading $b$ after preparing $a$.
Five hundred multinomial bootstrap samples repeat the calibration
inversion and nonlinear reconstruction.  The bars quantify shot and
assignment-estimator uncertainty, not correlated-preparation or
phase-averaging-model systematics.

This protocol measures initial polarizations and evaluates the
random-phase prediction under the product-weight assumption; it includes
no interacting evolution or Bell measurement at either size.
The 156-qubit estimate serves only as a model-assisted reference for the
predicted tilt dependence. Dynamical evidence comes from the time-resolved
six-qubit Bell measurements on \texttt{ibm\_boston}.

\subsection{Quasiperiodic kicked-XXZ chain realization}
\label{sec:supp_time_resolved_realization}

The microscopic interacting chain extends the comparison beyond the
directly encoded $\ell$-bit model. Its localized and ergodic magic
trajectories appear at $N=10$ in Fig.~\ref{fig:main_kicked_l10}, with the
$N=6$ comparison in Fig.~\ref{fig:supp_kicked_l6}.

\subsubsection{Floquet model and circuit construction}
\label{sec:supp_main_circuit}

Directly implementing the \(U(1)\)-symmetric random circuit used in
theory~\cite{SI-Xiao2026NonstabilizernessMpemba} is costly to realize on current hardware:
each charge-conserving two-qubit gate is drawn independently from a block-Haar
ensemble, so estimating the ensemble-averaged dynamics requires compiling and
executing many independent circuit realizations.  We instead retain the same
\(U(1)\) symmetry constraint in a single deterministic quasiperiodically
kicked XXZ drive, eliminating this ensemble overhead without changing the
symmetry setting. This is the drive shown in
Fig.~\ref{fig:hardware_workflow}(c).

The open quasiperiodically kicked XXZ model is described by the time-dependent
pulse Hamiltonian
\begin{equation}
H_{\mathrm{KXXZ}}(t)=\sum_{m\in\mathbb Z}\sum_{a=1}^{2}
\left[\delta(t-m-\tau_{a,\mathrm e})H_{\mathrm e,a}
+\delta(t-m-\tau_{a,\mathrm o})H_{\mathrm o,a}
+\delta(t-m-\tau_{a,\mathrm k})H_{\mathrm{kick},a}\right].
\label{eq:supp_kicked_hamiltonian}
\end{equation}
The six pulse times lie within one unit period and obey
\(\tau_{1,\mathrm e}<\tau_{1,\mathrm o}<\tau_{1,\mathrm k}
<\tau_{2,\mathrm e}<\tau_{2,\mathrm o}<\tau_{2,\mathrm k}\).  They only
specify the gate order.  With zero-based site labels, the two exchange pulses
are \(H_{\mu,a}=\frac{K_a}{2}
\sum_{(i,i+1)\in\mathcal B_\mu}
\left(X_iX_{i+1}+Y_iY_{i+1}\right)\), where
\(\mu\in\{\mathrm e,\mathrm o\}\).  Here
\(\mathcal B_{\mathrm e}=\{(0,1),(2,3),\ldots\}\) acts first and
\(\mathcal B_{\mathrm o}=\{(1,2),(3,4),\ldots\}\) acts second.  The
longitudinal kick contains both the Ising interaction and the quasiperiodic
field, \(H_{\mathrm{kick},a}=J_{z,a}\sum_{i=0}^{N-2}Z_iZ_{i+1}
+\sum_{i=0}^{N-1}h_i^{(a)}Z_i\), where
\(h_i^{(a)}=h_{0,a}+H_{\mathrm{qp},a}\cos(2\pi\beta i+\varphi)\) and
\(\beta=(\sqrt5-1)/2\).
Thus the anisotropic exchange is split between an \(XX+YY\) pulse and a
spatially quasiperiodic \(ZZ+Z\) kick, which gives the quasiperiodically
kicked XXZ construction
used throughout the quantum simulation.

The time-ordered evolution over one period is
\(U_{\mathrm F}=\mathcal T e^{-\ii\int_0^1H_{\mathrm{KXXZ}}(t)\dd t}
=U_2U_1\), where \(U_a=e^{-\ii H_{\mathrm{kick},a}}
e^{-\ii H_{\mathrm o,a}}e^{-\ii H_{\mathrm e,a}}\).
Because the bonds within either exchange set are disjoint and every term in
the kick is diagonal, each exponential factorizes exactly.  The four
elementary rotations and their CZ-based realizations are
\begingroup
\tikzcdset{arrows/.append style={line width=0.75pt}}
\tikzset{thickness/.style={line width=0.75pt}}
\begin{equation}
\resizebox{0.93\textwidth}{!}{$\displaystyle
\begin{aligned}
\text{\normalfont Exchange}\quad &
\left\{
\begin{aligned}
R_{XX,i,i+1}(K_a)&=e^{-\ii K_aX_iX_{i+1}/2}=
\vcenter{\hbox{\scalebox{0.82}{%
\begin{quantikz}[row sep=0.10cm,column sep=0.10cm]
\qw & \gate[2,style={fill=floquetExchange,draw=black!75,line width=0.75pt,rounded corners=2pt}]{R_{XX}(K_a)} & \qw\\
\qw & & \qw
\end{quantikz}}}}
=\vcenter{\hbox{\scalebox{0.82}{%
\begin{quantikz}[align equals at=1.5,column sep=0.12cm,row sep={0.60cm,between origins}]
\qw & \gate{H} & \ctrl{1} & \qw & \ctrl{1} & \gate{H} & \qw\\
\qw & \qw & \control{} & \gate{R_X(K_a)} & \control{} & \qw & \qw
\end{quantikz}}}} \\[0.5ex]
R_{YY,i,i+1}(K_a)&=e^{-\ii K_aY_iY_{i+1}/2}=
\vcenter{\hbox{\scalebox{0.82}{%
\begin{quantikz}[row sep=0.10cm,column sep=0.10cm]
\qw & \gate[2,style={fill=floquetExchange,draw=black!75,line width=0.75pt,rounded corners=2pt}]{R_{YY}(K_a)} & \qw\\
\qw & & \qw
\end{quantikz}}}}
=\vcenter{\hbox{\scalebox{0.82}{%
\begin{quantikz}[align equals at=1.5,column sep=0.12cm,row sep={0.60cm,between origins}]
\qw & \gate{R_X(\tfrac{\pi}{2})} & \qw & \ctrl{1} & \qw & \ctrl{1} & \qw & \gate{R_X(-\tfrac{\pi}{2})} & \qw\\
\qw & \gate{R_X(\tfrac{\pi}{2})} & \gate{H} & \control{} & \gate{R_X(K_a)} & \control{} & \gate{H} & \gate{R_X(-\tfrac{\pi}{2})} & \qw
\end{quantikz}}}}
\end{aligned}
\right. \\[1.0ex]
\text{\normalfont Kick}\quad &
\left\{
\begin{aligned}
R_{ZZ,i,i+1}(2J_{z,a})&=e^{-\ii J_{z,a}Z_iZ_{i+1}}=
\vcenter{\hbox{\scalebox{0.82}{%
\begin{quantikz}[row sep=0.10cm,column sep=0.10cm]
\qw & \gate[2,style={fill=floquetDiagonal,draw=black!75,line width=0.75pt,rounded corners=2pt}]{R_{ZZ}(2J_{z,a})} & \qw\\
\qw & & \qw
\end{quantikz}}}}
=\vcenter{\hbox{\scalebox{0.82}{%
\begin{quantikz}[align equals at=1.5,column sep=0.12cm,row sep={0.60cm,between origins}]
\qw & \qw & \ctrl{1} & \qw & \ctrl{1} & \qw & \qw\\
\qw & \gate{H} & \control{} & \gate{R_X(2J_{z,a})} & \control{} & \gate{H} & \qw
\end{quantikz}}}} \\[0.5ex]
R_{Z,i}(2h_i^{(a)})&=e^{-\ii h_i^{(a)}Z_i}=
\vcenter{\hbox{\scalebox{0.82}{%
\begin{quantikz}[column sep=0.10cm]
\qw & \gate[style={fill=floquetDiagonal,draw=black!75,line width=0.75pt,rounded corners=2pt}]{R_Z(2h_i^{(a)})} & \qw
\end{quantikz}}}}
\end{aligned}
\right.
\end{aligned}$}
\label{eq:supp_kicked_gate_definitions}
\end{equation}
\endgroup
The upper and lower wires represent sites $i$ and $i+1$, respectively;
each pair of connected filled dots is a native CZ gate.  Each displayed
two-qubit decomposition uses two CZ gates.  The $H$ and $R_X$ symbols
stand for single-qubit sequences, not additional native gate types.
In the $\{\mathrm{CZ},R_Z,SX,X\}$ basis used here,
$SX=e^{\ii\pi/4}R_X(\pi/2)$, and, up to global phases,
$H=R_Z(\pi/2)\,SX\,R_Z(\pi/2)$ and
$R_X(\alpha)=R_Z(\pi/2)\,SX\,R_Z(\alpha+\pi)\,SX\,R_Z(\pi/2)$.
The $R_Z$ rotations are virtual.  These identities specify the
elementary gate realizations; neighboring single-qubit gates can be
combined during compilation.

On the exchange bonds, the exact factorization is \(e^{-\ii H_{\mu,a}}=
\prod_{(i,i+1)\in\mathcal B_\mu}R_{XX,i,i+1}(K_a)
R_{YY,i,i+1}(K_a)\).  For the diagonal kick, it is
\(e^{-\ii H_{\mathrm{kick},a}}=
\prod_{i=0}^{N-1}R_{Z,i}(2h_i^{(a)})
\prod_{i=0}^{N-2}R_{ZZ,i,i+1}(2J_{z,a})\).
Four consecutive bulk sites of one substep are shown explicitly below.  The
half exchange gates at the upper and lower edges connect to neighboring sites
outside this window.  The complete Floquet period applies the same circuit
first with \(a=1\) and then with \(a=2\).

\begin{samepage}
\begin{equation}
\resizebox{0.88\textwidth}{!}{$\displaystyle
\vcenter{\hbox{\scalebox{1.35}{$\displaystyle U_a$}}}=
\vcenter{\hbox{\clipbox{0pt 0.65cm 0pt 0.65cm}{%
\begin{quantikz}[row sep={0.75cm,between origins},column sep=0.10cm]
\qw
  & \qw
  & \gate[2,style={fill=floquetBlock,draw=black!75,line width=0.75pt,rounded corners=2pt}]{\phantom{U}}
  & \qw \\
\qw
  & \gate[2,style={fill=floquetBlock,draw=black!75,line width=0.75pt,rounded corners=2pt}]{U}
  &
  & \qw \\
\qw
  &
  & \gate[2,style={fill=floquetBlock,draw=black!75,line width=0.75pt,rounded corners=2pt}]{U}
  & \qw \\
\qw
  & \gate[2,style={fill=floquetBlock,draw=black!75,line width=0.75pt,rounded corners=2pt}]{U}
  &
  & \qw \\
\qw
  &
  & \gate[2,style={fill=floquetBlock,draw=black!75,line width=0.75pt,rounded corners=2pt}]{\phantom{U}}
  & \qw
  \\
\qw
  & \qw
  &
  & \qw
\end{quantikz}}}}
=
\vcenter{\hbox{\clipbox{0pt 0.65cm 0pt 0.65cm}{%
\begin{quantikz}[row sep={0.75cm,between origins},column sep=0.08cm]
\qw
  & \qw
  & \gate[2,style={fill=floquetExchange,draw=black!75,line width=0.75pt,rounded corners=2pt}]{\phantom{R_{XX}(K_a)}}
  & \gate[2,style={fill=floquetExchange,draw=black!75,line width=0.75pt,rounded corners=2pt}]{\phantom{R_{YY}(K_a)}}
  & \qw
  & \qw
  & \qw
  & \qw
  & \qw
  & \gate[2,style={fill=floquetExchange,draw=black!75,line width=0.75pt,rounded corners=2pt}]{\phantom{R_{XX}(K_a)}}
  & \gate[2,style={fill=floquetExchange,draw=black!75,line width=0.75pt,rounded corners=2pt}]{\phantom{R_{YY}(K_a)}}
  & \qw
  & \qw
  & \qw
  & \qw \\
\qw
  & \qw
  &
  &
  & \gate[2,style={fill=floquetExchange,draw=black!75,line width=0.75pt,rounded corners=2pt}]{R_{XX}(K_a)}
  & \gate[2,style={fill=floquetExchange,draw=black!75,line width=0.75pt,rounded corners=2pt}]{R_{YY}(K_a)}
  & \qw
  & \gate[2,style={fill=floquetDiagonal,draw=black!75,line width=0.75pt,rounded corners=2pt}]{R_{ZZ}(2J_{z,a})}
  & \gate[style={fill=floquetDiagonal,draw=black!75,line width=0.75pt,rounded corners=2pt}]{R_Z(2h_0^{(a)})}
  &
  &
  & \gate[2,style={fill=floquetExchange,draw=black!75,line width=0.75pt,rounded corners=2pt}]{R_{XX}(K_a)}
  & \gate[2,style={fill=floquetExchange,draw=black!75,line width=0.75pt,rounded corners=2pt}]{R_{YY}(K_a)}
  & \qw
  & \qw \\
\qw
  & \qw
  & \gate[2,style={fill=floquetExchange,draw=black!75,line width=0.75pt,rounded corners=2pt}]{R_{XX}(K_a)}
  & \gate[2,style={fill=floquetExchange,draw=black!75,line width=0.75pt,rounded corners=2pt}]{R_{YY}(K_a)}
  &
  &
  & \gate[2,style={fill=floquetDiagonal,draw=black!75,line width=0.75pt,rounded corners=2pt}]{R_{ZZ}(2J_{z,a})}
  &
  & \gate[style={fill=floquetDiagonal,draw=black!75,line width=0.75pt,rounded corners=2pt}]{R_Z(2h_1^{(a)})}
  & \gate[2,style={fill=floquetExchange,draw=black!75,line width=0.75pt,rounded corners=2pt}]{R_{XX}(K_a)}
  & \gate[2,style={fill=floquetExchange,draw=black!75,line width=0.75pt,rounded corners=2pt}]{R_{YY}(K_a)}
  &
  &
  & \qw
  & \qw \\
\qw
  & \qw
  &
  &
  & \gate[2,style={fill=floquetExchange,draw=black!75,line width=0.75pt,rounded corners=2pt}]{R_{XX}(K_a)}
  & \gate[2,style={fill=floquetExchange,draw=black!75,line width=0.75pt,rounded corners=2pt}]{R_{YY}(K_a)}
  &
  & \gate[2,style={fill=floquetDiagonal,draw=black!75,line width=0.75pt,rounded corners=2pt}]{R_{ZZ}(2J_{z,a})}
  & \gate[style={fill=floquetDiagonal,draw=black!75,line width=0.75pt,rounded corners=2pt}]{R_Z(2h_2^{(a)})}
  &
  &
  & \gate[2,style={fill=floquetExchange,draw=black!75,line width=0.75pt,rounded corners=2pt}]{R_{XX}(K_a)}
  & \gate[2,style={fill=floquetExchange,draw=black!75,line width=0.75pt,rounded corners=2pt}]{R_{YY}(K_a)}
  & \qw
  & \qw \\
\qw
  & \qw
  & \gate[2,style={fill=floquetExchange,draw=black!75,line width=0.75pt,rounded corners=2pt}]{\phantom{R_{XX}(K_a)}}
  & \gate[2,style={fill=floquetExchange,draw=black!75,line width=0.75pt,rounded corners=2pt}]{\phantom{R_{YY}(K_a)}}
  &
  &
  & \qw
  &
  & \gate[style={fill=floquetDiagonal,draw=black!75,line width=0.75pt,rounded corners=2pt}]{R_Z(2h_3^{(a)})}
  & \gate[2,style={fill=floquetExchange,draw=black!75,line width=0.75pt,rounded corners=2pt}]{\phantom{R_{XX}(K_a)}}
  & \gate[2,style={fill=floquetExchange,draw=black!75,line width=0.75pt,rounded corners=2pt}]{\phantom{R_{YY}(K_a)}}
  &
  &
  & \qw
  & \qw \\
\qw
  & \qw
  &
  &
  & \qw
  & \qw
  & \qw
  & \qw
  & \qw
  &
  &
  & \qw
  & \qw
  & \qw
  & \qw
\end{quantikz}}}}.
$}
\label{eq:supp_kicked_substep_circuit}
\end{equation}
\end{samepage}

Every exchange gate conserves the magnetization on its two sites, while the
diagonal gates preserve each \(Z_i\) separately.  Hence
\([U_{\mathrm F},\sum_iZ_i]=0\) exactly, rather than only after a Trotter
limit.  Moving between the ergodic and localized parameter sets therefore
changes only the gate angles, not the symmetry or nearest-neighbor entangling
skeleton.  The localized set uses the optimized phason \(\varphi\),
whereas the ergodic control uses \(\varphi=0\).
Table~\ref{tab:supp_kicked_parameters} collects the circuit parameters; no
phason average is taken.  We diagnose the dynamical regimes by exact
diagonalization (ED) of \(U_{\mathrm F}\) in the largest \(U(1)\) sector.  For
the ordered quasienergies, let $\delta_n$ be the adjacent spacing, including
the cyclic wrap-around gap.  We compute
\(\langle r\rangle=\langle
\min(\delta_n,\delta_{n+1})/\max(\delta_n,\delta_{n+1})\rangle_n\).
At \(N=10\), ED gives
\(\langle r\rangle=0.428\) for the localized set, near the Poisson value
0.386, and 0.586 for the ergodic set, near the circular unitary ensemble (CUE)
value 0.600.  The ordered
drive has no antiunitary symmetry, making CUE the appropriate ergodic
reference.

\begin{table}[!htbp]
  \centering
  \small
  \setlength{\tabcolsep}{9pt}
  \renewcommand{\arraystretch}{1.08}
  \caption[Quasiperiodically kicked-XXZ circuit parameters.]{\textit{Quasiperiodically kicked-XXZ circuit parameters.}}
  \label{tab:supp_kicked_parameters}
  \begin{tabular*}{\textwidth}{@{\extracolsep{\fill}}lrr@{}}
  \toprule
  Quantity & $N=6$ & $N=10$ \\
  \midrule
  $\theta/\pi$ & 0.250, 0.310, 0.500 & 0.250, 0.350, 0.500 \\
  $(K_1,K_2)$ localized & (0.228, 0.284) & (0.201, 0.250) \\
  $(K_1,K_2)$ ergodic & (0.511, 0.637) & (0.511, 0.637) \\
  $(J_{z,1},J_{z,2})$ & (1.073, 0.827) & (0.495, 0.381) \\
  $H_{\mathrm{qp},1}=H_{\mathrm{qp},2}$ & 2.313 & 1.058 \\
  $(h_{0,1},h_{0,2})$ & (0.170, 0.290) & (0.170, 0.290) \\
  \bottomrule
\end{tabular*}
\end{table}

Alongside TFS, we use tilted N\'eel states (TNS) and tilted domain-wall
states (TDWS).  The three families are specified by base strings
\(s_i\in\{0,1\}\): TFS has \(s_i=0\), TNS has \(s_i=i\bmod2\), and TDWS has
\(s_i=0\) on the left half and \(s_i=1\) on the right half.  A base-0 site
receives \(R_y(\theta)\); a base-1 site receives
\(R_y(\pi-\theta)\).  Equivalently,
\begin{equation}
\begin{gathered}
\ket{\psi_0(\theta)}=
\bigotimes_{i=0}^{N-1}R_y\!\left[(1-s_i)\theta+s_i(\pi-\theta)\right]
\ket{\uparrow}, \\[0.8ex]
\begin{array}{@{}r@{\;}c@{\;}l@{}}
\mathrm{TFS}&=&
\vcenter{\hbox{\includegraphics[width=0.40\textwidth]{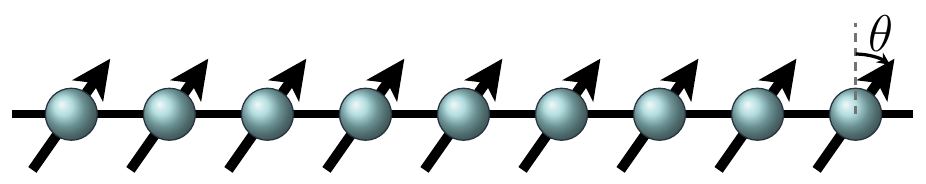}}} \\[0.7ex]
\mathrm{TNS}&=&
\vcenter{\hbox{\includegraphics[width=0.40\textwidth]{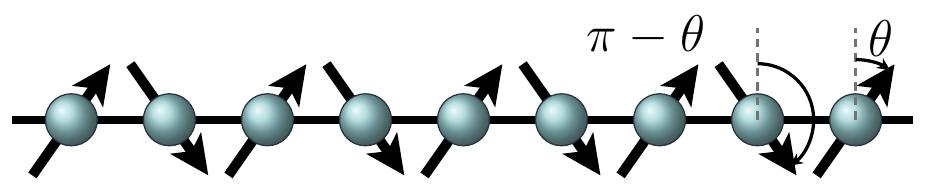}}} \\[0.7ex]
\mathrm{TDWS}&=&
\vcenter{\hbox{\includegraphics[width=0.40\textwidth]{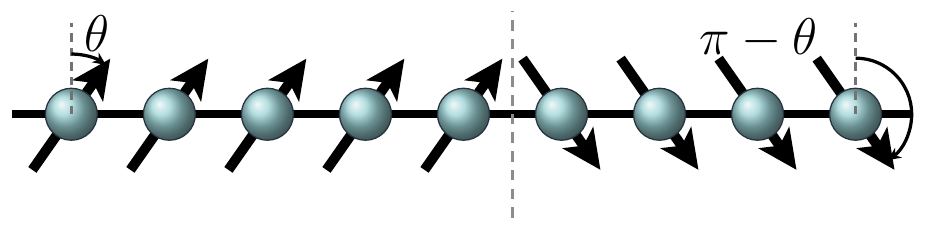}}}
\end{array}
\end{gathered}
\label{eq:supp_hardware_tilted_product_state}
\end{equation}
All three families therefore share the same symmetry-breaking magnitude
with distinct spatial patterns.  The interval
\(\pi/4\leq\theta\leq\pi/2\), rather than an empirical angle cut, follows
from the magic-ordering reversal derived in Sec.~\ref{sec:supp_tilt_interval}.
At \(\theta=\pi/2\), both local angles equal \(\pi/2\), so TFS, TNS, and TDWS
all reduce to \(\ket{+}^{\otimes N}\); this common dataset is shown in each
pattern panel but counted only once among the preparation targets.

\subsubsection{Target-state preparation and validation}
\label{sec:supp_approximate_quantum_compilation}

Each sampled state is prepared independently using target-state approximate
quantum compilation~\cite{SI-Khatri2019QuantumCompiling}, with
$\widetilde U(t,\theta)\ket{0}^{\otimes N}\simeq
U_{\mathrm F}^{t}\ket{\psi_0(\theta)}$ [Fig.~\ref{fig:hardware_workflow}(b)].
The optimization concerns this state, not the evolution operator on arbitrary
inputs. The circuit alternates one-qubit $U(2)$ and nearest-neighbor $U(4)$
layers, with depth selected per target. Twelve continuation steps,
$\lambda_r^{\mathrm{cont}}=r/12$ ($r=1,\ldots,12$), connect successive
integer times; bidirectional polar sweeps update the gate blocks.
Each $U(4)$ block decomposes into three native $\mathrm{CZ}$ gates and
one-qubit rotations on connected paths without SWAP routing.

Validation uses exact $N=6$ and 10 target states through the fidelity
$\mathcal F=|\bra{\psi(t,\theta)}\widetilde U(t,\theta)
\ket{0}^{\otimes N}|^2$ and the absolute 2-SRE bias.
Table~\ref{tab:supp_aqc_resources} reports these noiseless checks and gate
resources for 49 distinct $t>0$ targets per size and regime, counting the
common $\theta=\pi/2$ state once; $t=0$ requires only product-state preparation.
These checks bound the state-preparation contribution to the magic bias
in Figs.~\ref{fig:main_kicked_l10} and \ref{fig:supp_kicked_l6} before
hardware noise. Circuit depth measures preparation cost, not relaxation time.

\begin{table}[!htbp]
  \centering
  \footnotesize
  \setlength{\tabcolsep}{6pt}
  \renewcommand{\arraystretch}{1.06}
  \caption[Target-state preparation resources and validation.]{\textit{Target-state preparation resources and validation.}
  Median (range) over the 49 distinct $t>0$ targets for each size and regime,
  for the state-specific preparations $\widetilde U(t,\theta)$.}
  \label{tab:supp_aqc_resources}
  \begin{tabular*}{\textwidth}{@{\extracolsep{\fill}}llrrrr@{}}
  \toprule
  \multicolumn{2}{@{}l}{\multirow{2}{*}{Quantity}} & \multicolumn{2}{c}{$N=6$} & \multicolumn{2}{c}{$N=10$} \\
  \cmidrule(lr){3-4}\cmidrule(lr){5-6}
  & & Ergodic & Localized & Ergodic & Localized \\
  \midrule
  \multicolumn{2}{@{}l}{\# of logical $U(4)$ gates} & 10 (7--12) & 7 (5--10) & 81 (45--90) & 18 (18--36) \\
  \multicolumn{2}{@{}l}{Logical $U(4)$-gate depth} & 4 (3--5) & 3 (2--4) & 18 (10--20) & 4 (4--8) \\
  \multicolumn{2}{@{}l}{\# of native $\mathrm{CZ}$ gates} & 30 (21--36) & 21 (15--30) & 243 (135--270) & 54 (54--108) \\
  \multicolumn{2}{@{}l}{Native $\mathrm{CZ}$-gate depth} & 12 (9--15) & 9 (6--12) & 54 (30--60) & 12 (12--24) \\
  \midrule
  \multicolumn{2}{@{}l}{Min. fidelity} & 0.941 & 0.940 & 0.859 & 0.982 \\
  \multicolumn{2}{@{}l}{Max. $|\mathcal M_2|$ bias (bit)} & 0.143 & 0.148 & 0.147 & 0.090 \\
  \bottomrule
\end{tabular*}
\end{table}

\subsubsection{Randomized local-Pauli estimator}
\label{sec:supp_randomized_measurement}

The magic estimates in Figs.~\ref{fig:main_kicked_l10} and
\ref{fig:supp_kicked_l6} use randomized local-Pauli measurements.
Every shot begins with a freshly prepared state. A product basis
\(\bm b=(b_0,\ldots,b_{N-1})\), with \(b_i\in\{X,Y,Z\}\), is mapped to the
computational basis by local Clifford rotations and then measured.  Four
distinct single-copy shots are combined only in classical post-processing;
no four-copy circuit is executed~\cite{SI-Oliviero2022MeasuringMagic,SI-Huang2020ClassicalShadows,SI-Elben2023RandomizedMeasurementToolbox}.
For a possibly mixed hardware state, we
estimate the same fourth Pauli intensity,
\(\mathcal I_4(\rho)=2^{-N}\sum_P[\Tr(\rho P)]^4\), and the purity
\(\mathcal P(\rho)=\Tr\rho^2=2^{-N}\sum_P[\Tr(\rho P)]^2\), and report
the purity-normalized estimate
\(\mathcal M_2=-\log_2[\mathcal I_4(\rho)/\mathcal P(\rho)]\).
The denominator equals one for an
ideal pure state but reduces the leading mixedness bias in hardware data.

Finite random settings are balanced with randomized Latin blocks.  Each block
contains three settings; for every qubit, an independent permutation of
\(X,Y,Z\) assigns the three axes, so each axis occurs once per block while
the joint product bases remain random.  The same basis matrix is reused for
all tilts within one repeat, enabling paired comparisons without correlating
their shot noise.

For computational-basis outcome strings $\bm o_1,\ldots,\bm o_k$,
the local-Pauli exclusive-OR (XOR) kernel is
$K_k(\bm o_1,\ldots,\bm o_k)=2^N(-2)^{-|\bm o_1\oplus\cdots\oplus\bm o_k|}$,
where $|\cdot|$ denotes Hamming weight.
For a setting containing \(S\) shots, the order-four estimator uses all
\(\binom{S}{4}\) distinct quadruples and the purity estimator uses all
\(\binom{S}{2}\) distinct pairs~\cite{SI-Hoeffding1948UStatistics}.  A
Walsh--Hadamard transform representation of the XOR kernel and elementary symmetric
polynomials evaluate the complete sums in approximately \(O(S2^N)\) time,
rather than enumerating \(O(S^4)\) quadruples.  Within each repeat, moments
are averaged over settings before taking their ratio and logarithm; repeat
estimates are averaged only afterward.  This order prevents the nonlinear
logarithm from assigning disproportionate weight to a noisy setting.

\subsubsection{Readout correction and uncertainty}
\label{sec:supp_main_noise}

The per-qubit assignment matrix, calibrated from \(p_i(1|0)\) and
\(p_i(0|1)\), is inverted in tensor-product form and applied to each outcome
before the nonlinear U-statistic kernel, producing a signed
quasidistribution that is neither clipped nor projected onto the probability
simplex~\cite{SI-Bravyi2021Readout}.  Raw and mitigated estimates use the same
settings, shots, purity normalization, and aggregation; they differ only by
this per-shot inverse channel.  The correction targets factorized assignment
error and does not remove coherent, preparation, two-qubit, or
correlated-readout errors.

A paired hierarchical bootstrap~\cite{SI-Efron1979Bootstrap} resamples complete
three-setting Latin blocks and, when $R>1$, the $R$ independent acquisition
repeats. All tilts within a regime share the bootstrap draws, retaining
the paired comparisons used for the $\Delta\mathcal M_2$ insets in
Figs.~\ref{fig:main_kicked_l10} and \ref{fig:supp_kicked_l6}.
Bars are 95\% percentile intervals.  Localized $N=10$ has one acquisition repeat,
so its intervals capture Latin-block sampling fluctuations but not
between-acquisition variation.  Table~\ref{tab:supp_measurement_budgets}
gives the acquisition budgets.

\subsection{Devices and measurement budgets}
\label{sec:supp_device_properties}

Both dynamical schemes run on \texttt{ibm\_boston}, whereas
\texttt{ibm\_kingston} supplies the polarization data for the
model-assisted references in Fig.~\ref{fig:main_overview}(b).
Both processors contain 156 superconducting qubits on
the heavy-hex graph [Fig.~\ref{fig:supp_device_comparison}(a)].
Using the physical labels in Fig.~\ref{fig:supp_device_comparison}(a),
the kicked chains on \texttt{ibm\_boston} occupy
$(4,3,16,23,24,25)$ for $N=6$ and
$(4,3,16,23,24,25,37,45,44,43)$ for $N=10$, listed in logical-site order.
The six-qubit path is the first six sites of the ten-qubit path, preserving
the local hardware setting when changing size.
On \texttt{ibm\_kingston}, the six-qubit polarization reconstruction uses
$(141,148,91,69,10,50)$, whereas the $N=156$ reconstruction uses every
physical qubit, numbered $0$--$155$.
These reconstruction measurements require no connected path: they use
parallel one-qubit rotations
and readout without entangling gates.  Gate errors and coherence therefore
affect the two dynamical schemes, while the reconstruction of saturated magic is
limited primarily by preparation and assignment errors.

\begin{figure}[H]
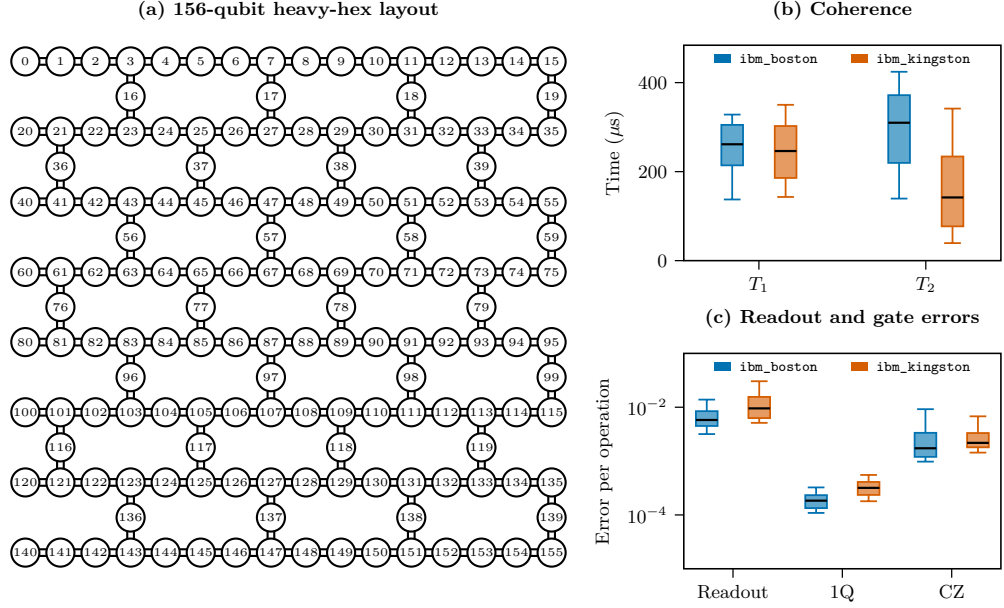

  \centering
  \insertdevicepropertiespdf
  \caption[Device properties.]{\textit{Device properties.}
  ($a$) The 156-qubit heavy-hex connectivity shared by both processors,
  with each site a qubit and each double line a calibrated
  nearest-neighbor bond.  ($b$) Distributions of the energy-relaxation
  time $T_1$ and transverse-coherence time $T_2$, and ($c$) of readout,
  single-qubit (1Q), and $\mathrm{CZ}$ error rates on a logarithmic
  axis, for \texttt{ibm\_boston} and \texttt{ibm\_kingston}.
  Boxes span the 25th to 75th percentiles, with medians marked and whiskers
  at the 10th and 90th percentiles, over all usable qubits in the frozen
  calibration snapshots.}
  \label{fig:supp_device_comparison}
\end{figure}

Table~\ref{tab:supp_device_summary} gives the corresponding 10th and
90th percentiles and medians.  Readout error is the per-qubit assignment
error; the 1Q entry is the calibrated identity-gate benchmark, and the
$\mathrm{CZ}$ entry is averaged over calibrated neighbors.  Blank
calibration entries are excluded from the affected observable.
The longer median $T_2$ and smaller $\mathrm{CZ}$ error of
\texttt{ibm\_boston} favor circuits containing entangling gates.
On \texttt{ibm\_kingston}, the shorter $T_2$ is less restrictive for
the rotation-and-readout protocol, and in-run all-up and all-down
calibrations correct the larger assignment error before reconstruction.

\begin{table}[H]
  \centering
  \scriptsize
  \setlength{\tabcolsep}{2.5pt}
  \renewcommand{\arraystretch}{1.08}
  \caption[Processor calibration summary.]{\textit{Processor calibration summary.}
  Here 2Q denotes two-qubit gates and Med. the median.
  Error units are given in the column headings.}
  \label{tab:supp_device_summary}
  \begin{tabular*}{\textwidth}{@{\extracolsep{\fill}}lccrrrrrrrrr@{}}
  \toprule
  \multirow{2}{*}{Device} & \multirow{2}{*}{Qubits} & \multirow{2}{*}{2Q gate} & \multicolumn{3}{c}{1Q gate error ($\times10^{-4}$)} & \multicolumn{3}{c}{2Q gate error ($\times10^{-3}$)} & \multicolumn{3}{c}{Readout error ($\times10^{-2}$)} \\
  \cmidrule(lr){4-6}\cmidrule(lr){7-9}\cmidrule(lr){10-12}
  & & & 10\% & 90\% & Med. & 10\% & 90\% & Med. & 10\% & 90\% & Med. \\
  \midrule
  \texttt{ibm\_boston} & 156 & $\mathrm{CZ}$ & 1.09 & 3.24 & 1.84 & 0.98 & 9.20 & 1.73 & 0.32 & 1.39 & 0.58 \\
  \texttt{ibm\_kingston} & 156 & $\mathrm{CZ}$ & 1.79 & 5.53 & 3.18 & 1.44 & 6.78 & 2.18 & 0.51 & 3.04 & 0.95 \\
  \bottomrule
\end{tabular*}
\par\vspace{0.55em}
\begin{tabular*}{\textwidth}{@{\extracolsep{\fill}}lrrrrrr@{}}
  \toprule
  \multirow{2}{*}{Device} & \multicolumn{3}{c}{$T_1$ ($\mu$s)} & \multicolumn{3}{c}{$T_2$ ($\mu$s)} \\
  \cmidrule(lr){2-4}\cmidrule(lr){5-7}
  & 10\% & 90\% & Med. & 10\% & 90\% & Med. \\
  \midrule
  \texttt{ibm\_boston} & 137 & 328 & 261 & 139 & 424 & 310 \\
  \texttt{ibm\_kingston} & 143 & 350 & 246 & 39 & 342 & 142 \\
  \bottomrule
\end{tabular*}
\end{table}

Table~\ref{tab:supp_measurement_budgets} collects the acquisition budgets.
The Bell protocol averages over disorder realizations, the local-Pauli
protocol samples measurement settings and acquisition repeats, and the
polarization-reconstruction protocol propagates count fluctuations through the polarization
reconstruction.  Their uncertainty conventions follow these different
sampling procedures: one disorder standard deviation for Bell trajectories,
95\% hierarchical-bootstrap intervals for kicked-XXZ trajectories, and
multinomial-bootstrap uncertainty for the reconstructed saturated value.
Calibration summaries alone do not quantify these errors or the bias of
the nonlinear magic estimators; the corrections are specified separately
within each implementation above.

\begin{table}[H]
  \centering
  \footnotesize
  \setlength{\tabcolsep}{5pt}
  \renewcommand{\arraystretch}{1.06}
  \caption[Measurement budgets.]{\textit{Measurement budgets.}
  $B$ is the number of Latin blocks and $S$ the shots per setting.
  $R$ counts acquisition repeats for the local-Pauli rows and disorder
  realizations for the Bell rows.  Bell noise scales refer to ZNE.
  Int., And., and calib. denote interacting, Anderson, and calibration.}
  \label{tab:supp_measurement_budgets}
  \begin{tabular*}{\textwidth}{@{\extracolsep{\fill}}lrrrrrr@{}}
  \toprule
  Protocol & $B$ & Settings & $S$ & $R$ & Preparations/state & Bootstrap \\
  \midrule
  \multicolumn{7}{@{}l@{}}{\emph{Kicked-XXZ one-copy local-Pauli measurement} (\texttt{ibm\_boston})} \\
  $N=6$, all panels & 20 & 60 & 128 & 6 & 46080 & 2000 \\
  $N=10$, localized & 200 & 600 & 384 & 1 & 230400 & 4000 \\
  $N=10$, ergodic & 200 & 600 & 512 & 8 & 2457600 & 4000 \\
  \midrule
  \multicolumn{7}{@{}l@{}}{\emph{$\ell$-bit two-copy Bell protocol} (\texttt{ibm\_boston})} \\
  Int., $t>0$ ($\lambda=1,2$) & \textemdash & 3 tilts $\times$ 5 times & 8000 & 15 & 240000 & \textemdash \\
  Int. $t=0$, And. ($\lambda=1$) & \textemdash & 3 tilts $\times$ ($1+6$) times & 8000 & 15 & 120000 & \textemdash \\
  \midrule
  \multicolumn{7}{@{}l@{}}{\emph{$\ell$-bit reconstruction of saturated magic} (\texttt{ibm\_kingston})} \\
  $N=6$ and $N=156$, 16 angles & \textemdash & 32 data $+$ 4 calib. & 4000 & 1 & 4000 & 500 \\
  \bottomrule
  \addlinespace[2pt]
  \multicolumn{7}{@{}l@{}}{\footnotesize Bell rows: $\lambda$ is the ZNE noise scale, and preparations per state $=$ (\# of scales) $\times\,S\times R$.} \\
\end{tabular*}
\end{table}

\section{Initial-state dependence of six-qubit magic Mpemba signatures}
\label{sec:supp_further_hardware_results}

Fig.~\ref{fig:supp_kicked_l6} presents the \(N=6\) kicked-XXZ magic
dynamics on \texttt{ibm\_boston}, complementing the \(N=10\) results in
Fig.~\ref{fig:main_kicked_l10}.  The six-qubit realization uses its own
prescreened Floquet parameters, compiled circuit set, sampling budget,
and tilt grid, as specified in Sec.~\ref{sec:supp_hardware_execution}.

In the ergodic regime, the magic-ordering reversal associated with the
magic Mpemba effect (MME) depends on the initial-state
pattern~\cite{SI-Xiao2026NonstabilizernessMpemba}: a reversal occurs for TFS
and TDWS, whereas TNS shows no such crossing.  In the localized regime,
all three initial-state families exhibit reversals and follow nearly
identical dynamics.  These \(N=6\) results reproduce the contrast observed
at \(N=10\), supporting a shared localized response across the initial
patterns studied.  Raw dots and mitigated
open-square symbols come from the same circuit records at identical time
coordinates; their difference isolates assignment correction.

\begin{figure}[!htbp]
  \centering
  \includegraphics[width=\textwidth]{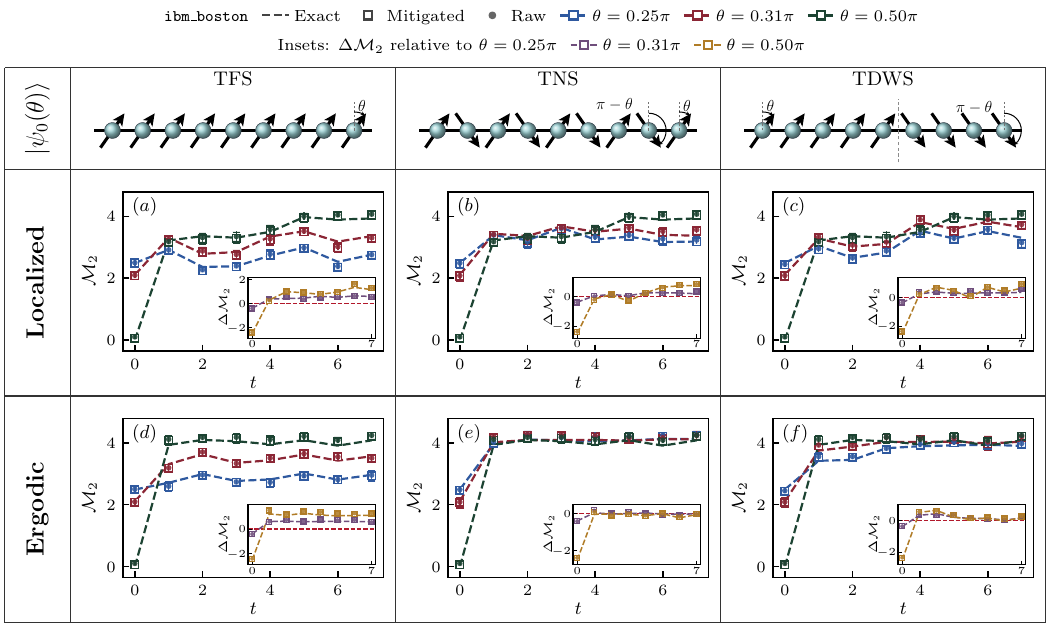}
  \caption[Kicked-XXZ magic dynamics on \texttt{ibm\_boston}.]{\textit{Kicked-XXZ magic dynamics on}
  \texttt{ibm\_boston}. As in Fig.~\ref{fig:main_kicked_l10} but for
  \(N=6\): top row ($a$--$c$), localized; bottom row ($d$--$f$), ergodic.
  Column headers illustrate TFS, TNS and TDWS. Colors label
  $\theta/\pi=0.25,0.31,0.50$; dashed lines are exact trajectories, filled
  circles raw estimates and open squares readout-mitigated estimates, with
  95\% hierarchical-bootstrap error bars. Insets show
  $\Delta\mathcal M_2(t,\theta)=\mathcal M_2(t,\theta)-\mathcal M_2(t,0.25\pi)$,
  with purple and gold denoting $\theta=0.31\pi$ and $0.50\pi$; the red
  dashed line marks zero.}
  \label{fig:supp_kicked_l6}
\end{figure}

\section{Theory of dephasing-induced magic-ordering reversal}
\label{sec:supp_lbit_theory}
\begingroup
\compactmathspacing

The effective Hamiltonian in Eq.~\eqref{eq:Hlbits} isolates how
interactions can relax magic while local memory persists. In the fully
many-body-localized regime, the
$\ell$-bit description expresses the Hamiltonian in terms of quasi-local
conserved pseudospins: their polarizations remain fixed, but their mutual
interactions generate dephasing~\cite{SI-Huse2014Phenomenology,SI-Serbyn2013LocalConservation}.
This separates the conserved configuration probabilities from the evolving
relative phases, allowing us to identify how each enters the Pauli moments
that determine magic.  We therefore use the effective Hamiltonian
\begin{equation}
 H_{\ell\text{-bit}}
 =\sum_i h_i\tau_i^z
 +\sum_{i<j}J_{ij}\tau_i^z\tau_j^z
 +\sum_{i<j<k}K_{ijk}\tau_i^z\tau_j^z\tau_k^z
 +\cdots.
 \label{eq:supp_lbit_theory_hamiltonian}
\end{equation}
Here $\tau_i^z$ are quasi-local conserved operators, and the interactions
decay exponentially with their spatial range.  All terms commute, so
evolution changes configuration phases while preserving their probabilities;
the interaction terms make the precession of each pseudospin depend on the
configuration of the others.  This provides a tractable setting for deriving
the initial and dephased magic orderings from the same conserved weights.

We work in the directly encoded basis $\tau_i^z=Z_i$ and retain terms
through three-body order in the realized and numerical $\ell$-bit models,
as in Eq.~\eqref{eq:supp_lbit_hardware_hamiltonian}.  This identification
is a modeling choice: in a microscopic localized spin chain, the conserved
$\ell$-bits are generally quasi-local dressed operators, and magic depends
on the choice of Pauli basis.  The derivation below therefore concerns the
directly encoded model; physical-spin dynamics is examined separately
in the microscopic models of Sec.~\ref{sec:supp_numerics}.

The analytical curves in Fig.~\ref{fig:main_overview}(b) predict opposite
initial and saturated tilt orderings. The processor trajectories in
Fig.~\ref{fig:main_overview}(c) show an interaction-induced crossing
that is absent in the Anderson control of Fig.~\ref{fig:endmatter_anderson}.
The derivations below determine the endpoint values, not the saturation
times. Earlier saturation from a greater initial distance is established
separately by exact numerical dynamics in End Matter
Fig.~\ref{fig:endmatter_localized_steady_state_time}, detailed in
Sec.~\ref{sec:supp_numerical_results}.
Fig.~\ref{fig:supp_theory_magic_ordering} summarizes the analytic angle
dependences derived below for (a) $\mathcal M_2$ and (b) $\Ba$,
including the finite-size phase-averaged saturated values of both measures.

For a Pauli string $P$, let
$p(P)=2^{-N}|\bra{\psi}P\ket{\psi}|^2$.  Purity gives
$\sum_Pp(P)=1$.  The fourth Pauli intensity is
\begin{equation}
 \Ifour(\psi)=2^{-N}\sum_{P\in\cP_N}|\bra{\psi}P\ket{\psi}|^4.
 \label{eq:supp_m2_moment}
\end{equation}
The 2-SRE is
\begin{equation}
 \mathcal M_2(\psi)=-\log_2\Ifour(\psi).
 \label{eq:supp_m2_definition}
\end{equation}
For comparison, the additive Bell magic is obtained from the eighth Pauli
moment~\cite{SI-ScalableMeasuresOfMagicResourceForQuantum,SI-Haug2024StabilizerEntropies},
\begin{equation}
 2^{-\Ba(\psi)}=2^{2N}\sum_{R,P\in\cP_N}p(R)^2p(P)p(PR),
 \label{eq:supp_bell_eighth_moment}
\end{equation}
where $PR$ is understood modulo the phase of the Pauli group.

\subsection{Initial magic ordering and the tilt interval for reversal}
\label{sec:supp_tilt_interval}

We use the same initial states and zero-based site labels as in the main
text.  In the directly encoded $\ell$-bit model, $\tau_i^z=Z_i$, and
\begin{equation}
 \ket{\psi_0(\theta)}=\bigotimes_{i=0}^{N-1}
 R_y\!\left[(1-s_i)\theta+s_i(\pi-\theta)\right]\ket{\uparrow}.
 \label{eq:supp_tilted_product_state}
\end{equation}
Here $\ket{\uparrow}=\ket{0}$ and $\eta_i=1-2s_i$.  TFS has $\eta_i=+1$
at every site, TNS has $\eta_i=(-1)^i$, and TDWS has $\eta_i=+1$ for
$0\leq i<N/2$ and $\eta_i=-1$ for $N/2\leq i<N$.  Direct substitution gives
the local Bloch vector
\begin{equation}
 \bigl(\langle X_i\rangle,\langle Y_i\rangle,\langle Z_i\rangle\bigr)^{\mathsf T}
 =\bigl(\sin\theta,0,\eta_i\cos\theta\bigr)^{\mathsf T}.
 \label{eq:supp_tilted_bloch_components}
\end{equation}
Thus the three patterns differ only in their longitudinal signs. At $t=0$ these signs
disappear from the even Pauli moments. In the independent-phase ensemble they
are removed by $z_i\mapsto\eta_i z_i$. The initial and phase-averaged saturated
values are therefore identical for TFS, TNS, and TDWS. These initial and
late-time values do not fix their finite-time trajectories or settling times; those are
assessed separately in Sec.~\ref{sec:supp_numerical_results}.

The symmetry-breaking strength, the initial moments, and the independent-phase
saturated value depend only on even powers of
$\sin\theta$ and $\cos\theta$.  Sign changes outside
$0\leq\theta\leq\pi/2$ therefore relabel the pattern or change only local
Pauli signs for the ordering problem considered here.  It is sufficient to
analyze this interval.  Because the state is a product state, the
variance of the conserved polarization is the sum of its local variances.
Direct substitution gives
\begin{align}
 \operatorname{Var}_{\theta}\!\left(\sum_i\tau_i^z\right)
 &=\sum_i\left[1-\eta_i^2\cos^2\theta\right],\nonumber\\
 &=N\sin^2\theta,\nonumber\\
 \frac{\partial}{\partial\theta}
 \operatorname{Var}_{\theta}\!\left(\sum_i\tau_i^z\right)
 &=N\sin(2\theta),\nonumber\\
 &\geq0.
 \label{eq:supp_u1_breaking_monotonicity}
\end{align}
The last inequality holds for $0\leq\theta\leq\pi/2$, so larger $\theta$
means stronger $U(1)$ breaking throughout this range.

The initial moments follow directly from the Bloch components in
Eq.~\eqref{eq:supp_tilted_bloch_components}.  The fourth intensity and its
logarithm obey
\begin{align}
 \Ifour(0,\theta)
 &=\left[1-\frac14\sin^2(2\theta)\right]^N,\nonumber\\
 \mathcal M_2(0,\theta)
 &=-N\log_2\left[1-\frac14\sin^2(2\theta)\right],\nonumber\\
 \frac{\mathcal M_2(0,\theta)}{N}
 &=-\log_2\left[1-\frac14\sin^2(2\theta)\right].
 \label{eq:supp_initial_m2}
\end{align}
The same substitution in the Bell moment gives
\begin{align}
 2^{-\Ba(0,\theta)}
 &=\left[1-\frac12\sin^2(2\theta)\right]^N,\nonumber\\
 \Ba(0,\theta)
 &=-N\log_2\left[1-\frac12\sin^2(2\theta)\right],\nonumber\\
 \frac{\Ba(0,\theta)}{N}
 &=-\log_2\left[1-\frac12\sin^2(2\theta)\right].
 \label{eq:supp_initial_ba}
\end{align}
We now differentiate rather than only quote the angular ordering.  For
$\mathcal M_2$,
\begin{align}
 \frac1N\frac{\partial \mathcal M_2(0,\theta)}{\partial\theta}
 &=-\frac{1}{\ln2}\frac{1}{1-\frac14\sin^2(2\theta)}
 \frac{\partial}{\partial\theta}\left[1-\frac14\sin^2(2\theta)\right]\nonumber\\
 &=\frac{\sin(4\theta)}{2\ln2[1-\frac14\sin^2(2\theta)]}.
 \label{eq:supp_initial_m2_monotonicity}
\end{align}
For $\Ba$,
\begin{align}
 \frac1N\frac{\partial\Ba(0,\theta)}{\partial\theta}
 &=-\frac{1}{\ln2}\frac{1}{1-\frac12\sin^2(2\theta)}
 \frac{\partial}{\partial\theta}\left[1-\frac12\sin^2(2\theta)\right]\nonumber\\
 &=\frac{\sin(4\theta)}{\ln2[1-\frac12\sin^2(2\theta)]}.
 \label{eq:supp_initial_ba_monotonicity}
\end{align}
Both derivatives are positive below $\pi/4$ and negative above $\pi/4$.
Hence $\pi/4\leq\theta\leq\pi/2$ is the only branch on which a larger tilt
simultaneously means stronger $U(1)$ breaking and smaller initial magic.  The
interacting saturated value derived below has the opposite angular ordering in the
thermodynamic limit, identifying a candidate interval for magic Mpemba
dynamics. Saturation-time ordering is determined numerically in
Sec.~\ref{sec:supp_numerical_results} and End Matter
Fig.~\ref{fig:endmatter_localized_steady_state_time}.
This is the orange-shaded interval in
Fig.~\ref{fig:main_overview}(b) and motivates the tilt range used in the main
quantum simulation.

\subsection{Magic dynamics under noninteracting Anderson localization}
\label{sec:supp_local_precession}

The noninteracting control in Fig.~\ref{fig:endmatter_anderson} removes
every $\ell$-bit interaction, leaving
\begin{equation}
 H_{\mathrm{And}}=\sum_i h_i\tau_i^z.
 \label{eq:supp_anderson_hamiltonian}
\end{equation}
Each spin only precesses about its local $z$ axis.  Its longitudinal component
and its two transverse components are obtained by a rotation through $2h_it$:
\begin{align}
 \langle X_i(t)\rangle
 &=\sin\theta\cos(2h_it),\nonumber\\
 \langle Y_i(t)\rangle
 &=\sin\theta\sin(2h_it),\nonumber\\
 \langle Z_i(t)\rangle
 &=\eta_i\cos\theta.
 \label{eq:supp_anderson_bloch_vector}
\end{align}
The state therefore remains a product state.  Substituting these three
components into the fourth Pauli intensity gives
\begin{align}
 \Ifour(t,\theta)
 &=\prod_i\frac{1+\langle X_i(t)\rangle^4
 +\langle Y_i(t)\rangle^4+\langle Z_i(t)\rangle^4}{2}\nonumber\\
 &=\prod_i\frac{1+\cos^4\theta
 +\sin^4\theta[\cos^4(2h_it)+\sin^4(2h_it)]}{2}\nonumber\\
 &=\prod_i\frac{1+\cos^4\theta+\sin^4\theta
 -\frac12\sin^4\theta\sin^2(4h_it)}{2}\nonumber\\
 &=\prod_i\left[1-\frac14\sin^2(2\theta)
 -\frac14\sin^4\theta\sin^2(4h_it)\right].
 \label{eq:supp_anderson_i4_exact}
\end{align}
Here we used $\cos^4x+\sin^4x=1-\sin^2(2x)/2$ in the third line and
$\sin^4\theta+\cos^4\theta=1-\sin^2(2\theta)/2$ in the last line.

The Bell moment also factorizes site by site.  For each pure single-spin
factor, direct evaluation of the Bell commutation character gives the fourth
powers of the three Bloch components and one mixed sixth-order term.  Inserting
Eq.~\eqref{eq:supp_anderson_bloch_vector} then yields
\begin{align}
 2^{-\Ba(t,\theta)}
 &=\prod_i\left[\langle X_i(t)\rangle^4
 +\langle Y_i(t)\rangle^4+\langle Z_i(t)\rangle^4
 +2\langle X_i(t)\rangle^2\langle Y_i(t)\rangle^2
 \langle Z_i(t)\rangle^2\right]\nonumber\\
 &=\prod_i\left\{\sin^4\theta[\cos^4(2h_it)+\sin^4(2h_it)]
 +\cos^4\theta
 +2\sin^4\theta\cos^2\theta\cos^2(2h_it)\sin^2(2h_it)\right\}\nonumber\\
 &=\prod_i\left[\sin^4\theta+\cos^4\theta
 -\frac12\sin^4\theta(1-\cos^2\theta)\sin^2(4h_it)\right]\nonumber\\
 &=\prod_i\left[1-\frac12\sin^2(2\theta)
 -\frac12\sin^6\theta\sin^2(4h_it)\right].
 \label{eq:supp_anderson_bell_exact}
\end{align}

The Anderson Hamiltonian has no pointwise steady state: the exact expressions
above remain quasiperiodic.  A field average damps the local oscillations, but
does not generate an interaction-controlled relaxation scale.  To see this
explicitly, rewrite the time-dependent term in the fourth moment as
\begin{equation}
 -\frac14\sin^4\theta\sin^2(4ht)
 =-\frac18\sin^4\theta+\frac18\sin^4\theta\cos(8ht).
 \label{eq:supp_anderson_i4_oscillation}
\end{equation}
The corresponding Bell term is
\begin{equation}
 -\frac12\sin^6\theta\sin^2(4ht)
 =-\frac14\sin^6\theta+\frac14\sin^6\theta\cos(8ht).
 \label{eq:supp_anderson_bell_oscillation}
\end{equation}
After averaging over the fields and dividing each oscillating part by its
initial amplitude, the two normalized expressions become
$\overline{\cos(8ht)}$.
The tilt cancels completely from both normalized kernels.  It changes only the
amplitude of the local precession, not its decay law.  For the tilt angles used
in our Anderson comparison, Eqs.~\eqref{eq:supp_anderson_i4_exact} and
\eqref{eq:supp_anderson_bell_exact} preserve the ordering of the trajectories
throughout the observed evolution.  This is the analytic counterpart of the
Anderson reference measured in Fig.~\ref{fig:endmatter_anderson}: the selected
trajectories do not exhibit a magic-ordering reversal.

The blue solid curve for the Anderson saturated value in Fig.~\ref{fig:supp_theory_magic_ordering}(a)
is obtained by independent uniform averaging of the local precession phases. With
$\overline{\sin^2(4h_i t)}=1/2$, Eq.~\eqref{eq:supp_anderson_i4_exact}
gives $\overline{\Ifour}=[1-\tfrac14\sin^2(2\theta)-\tfrac18\sin^4\theta]^N$.
The plotted density is therefore
$-N^{-1}\log_2\overline{\Ifour}=-\log_2[1-\tfrac14\sin^2(2\theta)-\tfrac18\sin^4\theta]$.
Applying the same local-phase average to
Eq.~\eqref{eq:supp_anderson_bell_exact} gives
\begin{equation}
 -\frac1N\log_2\overline{2^{-\Ba}}
 =-\log_2\left[1-\frac12\sin^2(2\theta)-\frac14\sin^6\theta\right],
 \label{eq:supp_anderson_ba_phase_density}
\end{equation}
which is the blue curve in Fig.~\ref{fig:supp_theory_magic_ordering}(b).
Both panels use the same moment-before-logarithm convention as the interacting
saturated values below, but these Anderson curves average independent local phases
rather than independent configuration phases. They are not pointwise
long-time limits of a finite chain.

\subsection{Phase-averaged saturated magic and interaction-induced magic-ordering reversal}
\label{sec:supp_farp}

For evolution under the three-body truncation of
Eq.~\eqref{eq:supp_lbit_theory_hamiltonian}, each configuration probability
remains fixed.  The tilted initial states give the weight
\begin{align}
 w(\bm z)
 &=|\braket{\bm z|\psi(t,\theta)}|^2,\nonumber\\
 &=\prod_i\frac{1+\eta_i z_i\cos\theta}{2}.
 \label{eq:supp_frozen_amplitudes}
\end{align}
Time evolution accumulates phases $\phi_t(\bm z)=-E(\bm z)t$, where
$E(\bm z)$ is the configuration energy, but never transports these
probabilities. The broad set of interaction scales motivates a random-phase
approximation to the coarse-grained late-time moments. In this approximation,
each accumulated configuration phase is represented by an independent
uniform phase, giving
\begin{equation}
 \ket{\psi_{\mathrm{rp}}(\theta;\phi)}
 =\sum_{\bm z}\sqrt{w(\bm z)}e^{\ii\phi(\bm z)}\ket{\bm z}.
 \label{eq:supp_random_phase_state}
\end{equation}
This frozen-amplitude random-phase ensemble defines the saturated-magic
prediction in Fig.~\ref{fig:main_overview}(b).
Throughout this subsection, the overbar means an independent average over the
configuration phases applied to the moment before taking its logarithm.  The
result is an annealed, coarse-grained saturated value; a finite closed realization
remains quasiperiodic and has no pointwise $t\rightarrow\infty$ limit.
As in the main text, we write
$\mathcal M_2^{\mathrm{sat}}(\theta)=-\log_2\overline{\Ifour(\theta)}$;
this is not the phase average of $\mathcal M_2$ itself.

We now calculate its fourth Pauli moment instead of quoting it.  Up to an
irrelevant Hermitian phase, write a Pauli string as
$P=X^{\bm u}Z^{\bm v}$, where $u_i,v_i\in\{0,1\}$.  For the Ising-valued
configuration $z_i=\pm1$, the notation $\bm z\mathbin\oplus\bm u$ means that
site $i$ is replaced by $(-1)^{u_i}z_i$.  Substitution of
Eq.~\eqref{eq:supp_random_phase_state} gives
\begin{equation}
 \langle P\rangle
 =\sum_{\bm z}\left(\prod_i z_i^{v_i}\right)
 \sqrt{w(\bm z)w(\bm z\mathbin\oplus\bm u)}
 e^{\ii[\phi(\bm z\mathbin\oplus\bm u)-\phi(\bm z)]}.
 \label{eq:supp_random_phase_pauli_expectation}
\end{equation}
First consider a diagonal string, for which no spin is flipped.  All random
phases then cancel, and the expectation value factorizes as
\begin{equation}
 \langle Z^{\bm v}\rangle
 =\prod_i(\eta_i\cos\theta)^{v_i}.
 \label{eq:supp_diagonal_pauli_expectation}
\end{equation}
Summing its fourth power over every diagonal string gives
\begin{align}
 2^{-N}\sum_{\bm v}|\langle Z^{\bm v}\rangle|^4
 &=2^{-N}\sum_{\bm v}
 \prod_i\cos^{4v_i}\theta\nonumber\\
 &=2^{-N}(1+\cos^4\theta)^N\nonumber\\
 &=\left(\frac{1+\cos^4\theta}{2}\right)^N.
 \label{eq:supp_diagonal_pauli_fourth_moment}
\end{align}
The second line follows by choosing independently at each site whether the
diagonal string contains the identity or $Z$.

Now let the string flip at least one spin.  The configurations in
Eq.~\eqref{eq:supp_random_phase_pauli_expectation} form pairs
$\{\bm z,\bm z\mathbin\oplus\bm u\}$.  Number these pairs by $r$ and choose
one representative $\bm z_r$ from each pair.  The two conjugate phase terms
in pair $r$ combine into a cosine with a uniform phase $\varphi_r$.  Direct
integration of its second and fourth powers gives
\begin{align}
 \overline{\left|2\sqrt{w(\bm z_r)w(\bm z_r\mathbin\oplus\bm u)}
 \cos\varphi_r\right|^2}
 &=2w(\bm z_r)w(\bm z_r\mathbin\oplus\bm u),\nonumber\\
 \overline{\left|2\sqrt{w(\bm z_r)w(\bm z_r\mathbin\oplus\bm u)}
 \cos\varphi_r\right|^4}
 &=6w^2(\bm z_r)w^2(\bm z_r\mathbin\oplus\bm u).
 \label{eq:supp_single_pair_phase_moments}
\end{align}
Different pairs have independent phases and zero mean.  Hence every term in
the fourth-power expansion vanishes unless each pair index appears an even
number of times.  Keeping the four-equal-index terms and the two-pair terms
gives
\begin{align}
 \overline{|\langle X^{\bm u}Z^{\bm v}\rangle|^4}
 &=6\sum_r w^2(\bm z_r)w^2(\bm z_r\mathbin\oplus\bm u)+24\sum_{r<s}w(\bm z_r)w(\bm z_r\mathbin\oplus\bm u)
 w(\bm z_s)w(\bm z_s\mathbin\oplus\bm u).
 \label{eq:supp_pairwise_phase_expansion}
\end{align}
The unrestricted sum over $\bm z$ counts each pair twice.  Therefore the last
line can be rearranged without any pair notation:
\begin{align}
 \sum_{\bm z}w(\bm z)w(\bm z\mathbin\oplus\bm u)
 &=2\sum_r w(\bm z_r)w(\bm z_r\mathbin\oplus\bm u),\nonumber\\
 \sum_{\bm z}w^2(\bm z)w^2(\bm z\mathbin\oplus\bm u)
 &=2\sum_r w^2(\bm z_r)w^2(\bm z_r\mathbin\oplus\bm u),\nonumber\\
 \overline{|\langle X^{\bm u}Z^{\bm v}\rangle|^4}
 &=3\left[\sum_{\bm z}w(\bm z)w(\bm z\mathbin\oplus\bm u)\right]^2
 -3\sum_{\bm z}w^2(\bm z)w^2(\bm z\mathbin\oplus\bm u).
 \label{eq:supp_pairwise_phase_average}
\end{align}
The final expression is independent of $\bm v$.  It remains to evaluate the
two configuration sums by inserting the product weight in
Eq.~\eqref{eq:supp_frozen_amplitudes}.

For a site that is not flipped, the local factor in the first configuration
sum is
\begin{align}
 \sum_{z_i}\left(\frac{1+\eta_i z_i\cos\theta}{2}\right)^2
 &=\frac{(1+\cos\theta)^2+(1-\cos\theta)^2}{4}\nonumber\\
 &=\frac{1+\cos^2\theta}{2}.
 \label{eq:supp_unflipped_first_convolution}
\end{align}
For a flipped site, the corresponding factor is
\begin{align}
 \sum_{z_i}\frac{1+\eta_i z_i\cos\theta}{2}
 \frac{1-\eta_i z_i\cos\theta}{2}
 &=\frac{2(1-\cos^2\theta)}{4}\nonumber\\
 &=\frac{1-\cos^2\theta}{2}.
 \label{eq:supp_flipped_first_convolution}
\end{align}
Multiplying these local results yields
\begin{equation}
 \sum_{\bm z}w(\bm z)w(\bm z\mathbin\oplus\bm u)
 =\prod_i\left(\frac{1+\cos^2\theta}{2}\right)^{1-u_i}
 \left(\frac{1-\cos^2\theta}{2}\right)^{u_i}.
 \label{eq:supp_first_configuration_convolution}
\end{equation}

The squared-weight convolution is calculated in the same way, but the local
powers are now fourth order.  At an unflipped site,
\begin{align}
 \sum_{z_i}\left(\frac{1+\eta_i z_i\cos\theta}{2}\right)^4
 &=\frac{(1+\cos\theta)^4+(1-\cos\theta)^4}{16}\nonumber\\
 &=\frac{1+6\cos^2\theta+\cos^4\theta}{8}.
 \label{eq:supp_unflipped_second_convolution}
\end{align}
At a flipped site,
\begin{align}
 \sum_{z_i}\left[
 \frac{(1+\eta_i z_i\cos\theta)(1-\eta_i z_i\cos\theta)}{4}
 \right]^2
 &=\frac{2(1-\cos^2\theta)^2}{16}\nonumber\\
 &=\frac{(1-\cos^2\theta)^2}{8}.
 \label{eq:supp_flipped_second_convolution}
\end{align}
Therefore
\begin{equation}
 \sum_{\bm z}w^2(\bm z)w^2(\bm z\mathbin\oplus\bm u)
 =\prod_i\left(\frac{1+6\cos^2\theta+\cos^4\theta}{8}\right)^{1-u_i}
 \left(\frac{(1-\cos^2\theta)^2}{8}\right)^{u_i}.
 \label{eq:supp_second_configuration_convolution}
\end{equation}

We can now perform the sum over every flip pattern $\bm u$.  Applying the
binomial theorem to the square of
Eq.~\eqref{eq:supp_first_configuration_convolution} gives
\begin{align}
 \sum_{\bm u}\left[\sum_{\bm z}w(\bm z)
 w(\bm z\mathbin\oplus\bm u)\right]^2
 &=\left[\left(\frac{1+\cos^2\theta}{2}\right)^2
 +\left(\frac{1-\cos^2\theta}{2}\right)^2\right]^N\nonumber\\
 &=\left(\frac{1+\cos^4\theta}{2}\right)^N.
 \label{eq:supp_sum_first_convolution}
\end{align}
For the unflipped pattern alone, the same squared convolution is
\begin{equation}
 \left[\sum_{\bm z}w^2(\bm z)\right]^2
 =\left(\frac{(1+\cos^2\theta)^2}{4}\right)^N.
 \label{eq:supp_zero_flip_first_convolution}
\end{equation}
Likewise, summing Eq.~\eqref{eq:supp_second_configuration_convolution} over all
flip patterns gives
\begin{align}
 \sum_{\bm u,\bm z}w^2(\bm z)w^2(\bm z\mathbin\oplus\bm u)
 &=\left[\frac{1+6\cos^2\theta+\cos^4\theta}{8}
 +\frac{(1-\cos^2\theta)^2}{8}\right]^N\nonumber\\
 &=\left(\frac{(1+\cos^2\theta)^2}{4}\right)^N.
 \label{eq:supp_sum_second_convolution}
\end{align}
For the unflipped pattern alone,
\begin{equation}
 \sum_{\bm z}w^4(\bm z)
 =\left(\frac{1+6\cos^2\theta+\cos^4\theta}{8}\right)^N.
 \label{eq:supp_zero_flip_second_convolution}
\end{equation}

We now assemble the result without skipping the subtraction of the zero-flip
pattern.  The diagonal contribution is
Eq.~\eqref{eq:supp_diagonal_pauli_fourth_moment}.  For every nonzero
$\bm u$, Eq.~\eqref{eq:supp_pairwise_phase_average} is independent of
$\bm v$.  There are $2^N$ choices of $\bm v$, which cancel the prefactor
$2^{-N}$ in $\Ifour$.  Thus
\begin{align}
 \overline{\Ifour(\theta)}
 &=\left(\frac{1+\cos^4\theta}{2}\right)^N
 +3\sum_{\bm u\neq\bm0}\left[\sum_{\bm z}w(\bm z)
 w(\bm z\mathbin\oplus\bm u)\right]^2-3\sum_{\bm u\neq\bm0}\sum_{\bm z}
 w^2(\bm z)w^2(\bm z\mathbin\oplus\bm u).
 \label{eq:supp_random_phase_i4_before_subtraction}
\end{align}
The first nonzero-flip sum equals the sum over all $\bm u$ minus the
$\bm u=\bm0$ term.  Substituting
Eqs.~\eqref{eq:supp_sum_first_convolution} and
\eqref{eq:supp_zero_flip_first_convolution} gives
\begin{align}
 \sum_{\bm u\neq\bm0}\left[\sum_{\bm z}w(\bm z)
 w(\bm z\mathbin\oplus\bm u)\right]^2
 &=\sum_{\bm u}\left[\sum_{\bm z}w(\bm z)
 w(\bm z\mathbin\oplus\bm u)\right]^2
 -\left[\sum_{\bm z}w^2(\bm z)\right]^2\nonumber\\
 &=\left(\frac{1+\cos^4\theta}{2}\right)^N
 -\left(\frac{(1+\cos^2\theta)^2}{4}\right)^N.
 \label{eq:supp_nonzero_first_convolution}
\end{align}
The second nonzero-flip sum is obtained in exactly the same way, now using
Eqs.~\eqref{eq:supp_sum_second_convolution} and
\eqref{eq:supp_zero_flip_second_convolution}:
\begin{align}
 \sum_{\bm u\neq\bm0}\sum_{\bm z}
 w^2(\bm z)w^2(\bm z\mathbin\oplus\bm u)
 &=\sum_{\bm u}\sum_{\bm z}w^2(\bm z)w^2(\bm z\mathbin\oplus\bm u)
 -\sum_{\bm z}w^4(\bm z)\nonumber\\
 &=\left(\frac{(1+\cos^2\theta)^2}{4}\right)^N
 -\left(\frac{1+6\cos^2\theta+\cos^4\theta}{8}\right)^N.
 \label{eq:supp_nonzero_second_convolution}
\end{align}
Inserting these two differences into
Eq.~\eqref{eq:supp_random_phase_i4_before_subtraction} gives every numerical
coefficient explicitly:
\begin{align}
 \overline{\Ifour(\theta)}
 &=\left(\tfrac{1+\cos^4\theta}{2}\right)^N
 +3\left[\left(\tfrac{1+\cos^4\theta}{2}\right)^N
 -\left(\tfrac{(1+\cos^2\theta)^2}{4}\right)^N\right]-3\left[\left(\tfrac{(1+\cos^2\theta)^2}{4}\right)^N
 -\left(\tfrac{1+6\cos^2\theta+\cos^4\theta}{8}\right)^N\right]\nonumber\\
 &=\left(1+3\right)\left(\frac{1+\cos^4\theta}{2}\right)^N
 +\left(-3-3\right)\left(\frac{(1+\cos^2\theta)^2}{4}\right)^N
 +3\left(\frac{1+6\cos^2\theta+\cos^4\theta}{8}\right)^N\nonumber\\
 &=4\bigl[(1+\cos^4\theta)/2\bigr]^N-6\bigl[(1+\cos^2\theta)^2/4\bigr]^N+3\bigl[(1+6\cos^2\theta+\cos^4\theta)/8\bigr]^N.
 \label{eq:supp_random_phase_i4_exact}
\end{align}
Eq.~\eqref{eq:supp_random_phase_i4_exact} generates the finite-size
late-time curves in Figs.~\ref{fig:main_overview}(b) and
\ref{fig:supp_theory_magic_ordering}(a), as well as the
homogeneous limit of the full-register reconstruction in
Sec.~\ref{sec:supp_lbit_full_register}.  Its second and third terms are finite-size
corrections. Fig.~\ref{fig:supp_theory_magic_ordering}(a) compares
$N=6,10,14,20,156$ with the thermodynamic limit, showing how the saturated value
density approaches the large-$N$ curve. The finite-size terms can produce a
shallow turnover close to $\theta=\pi/2$ at
small $N$, so the exact finite-size ordering must be checked at the compared
angles.  The thermodynamic ordering follows from the leading term derived
next.
Using the main-text notation $A=(1+\cos^4\theta)/2$, the leading term is
$4A^N$.  For fixed nonzero tilt, it has the largest exponential base.
Dividing by this leading base makes the dominance explicit:
\begin{align}
 \frac{\overline{\Ifour(\theta)}}
 {[(1+\cos^4\theta)/2]^N}
 &=4-6\left[\frac{(1+\cos^2\theta)^2}
 {2(1+\cos^4\theta)}\right]^N
 +3\left[\frac{1+6\cos^2\theta+\cos^4\theta}
 {4(1+\cos^4\theta)}\right]^N.
 \label{eq:supp_random_phase_i4_leading_base}
\end{align}
Both ratios raised to the $N$-th power are smaller than one for fixed nonzero
tilt.  Taking the logarithm and dividing by $N$ therefore gives
\begin{align}
 \lim_{N\rightarrow\infty}
 \frac{\mathcal M_2^{\mathrm{sat}}(\theta)}{N}
 &=-\log_2\left(\frac{1+\cos^4\theta}{2}\right)\nonumber\\
 &=1-\log_2(1+\cos^4\theta).
 \label{eq:supp_random_phase_m2_density}
\end{align}
This is the density obtained from the annealed fourth intensity.  Its
identification with a typical long-time-window magic density assumes
self-averaging of the random-phase ensemble; the finite-size expression is
kept explicitly when comparing with Fig.~\ref{fig:main_overview}(b).

We next calculate the Bell expression.  Let $q(R)$ be the probability that two
independently sampled Pauli strings differ by $R$.  Direct substitution gives
\begin{equation}
 q(R)=\sum_P p(P)p(PR).
 \label{eq:supp_bell_difference_probability}
\end{equation}
The commutation character of two sampled differences gives
\begin{equation}
 2^{-\Ba}=\sum_{R,S}q(R)q(S)(-1)^{[R,S]}.
 \label{eq:supp_bell_commutation_character}
\end{equation}
For a pure state, the symplectic Fourier transform of the Pauli probability is
\begin{equation}
 \sum_Sp(S)(-1)^{[R,S]}=2^Np(R).
 \label{eq:supp_pure_pauli_fourier_identity}
\end{equation}
We insert Eq.~\eqref{eq:supp_bell_difference_probability} into the Fourier
transform of $q(S)$ and change the summation variable from $S$ to $PS$:
\begin{align}
 \sum_Sq(S)(-1)^{[R,S]}
 &=\sum_{S,P}p(P)p(PS)(-1)^{[R,S]}\nonumber\\
 &=\left[\sum_Pp(P)(-1)^{[R,P]}\right]^2\nonumber\\
 &=2^{2N}p(R)^2.
 \label{eq:supp_bell_difference_fourier_transform}
\end{align}
Substitution into Eq.~\eqref{eq:supp_bell_commutation_character} then gives
\begin{align}
 2^{-\Ba}
 &=2^{2N}\sum_Rq(R)p(R)^2\nonumber\\
 &=2^{2N}\sum_{R,P}p(R)^2p(P)p(PR)\nonumber\\
 &=2^{-2N}\sum_{R,P}|\langle R\rangle|^4
 |\langle P\rangle|^2|\langle PR\rangle|^2.
 \label{eq:supp_bell_eighth_moment_expanded}
\end{align}
The last line follows by inserting
$p(P)=2^{-N}|\langle P\rangle|^2$ four times. Expanding the eight
expectation-value factors produces sixteen wave-function amplitudes:
eight carry a positive configuration phase and eight a negative one.
An independent-phase average retains precisely those terms for which each
configuration occurs equally often with both signs.

We evaluate this phase average with all finite-size contributions retained
and plot $-N^{-1}\log_2\overline{2^{-\Ba}}$ for $N=6,10,14,20,156$
in Fig.~\ref{fig:supp_theory_magic_ordering}(b). The result is an
independent-phase ensemble average, rather than a pointwise late-time
limit of a finite Hamiltonian. We checked the calculation against complete
phase quadrature for $N=1,2$ and direct random-phase sampling at $N=3$.

The contraction with the largest single-site weight consists of two copies of
the diagonal fourth-moment contraction.  Its single-site substitution is
\begin{align}
 \frac14(1+\cos^4\theta+\cos^4\theta+\cos^8\theta)
 &=\frac14(1+\cos^4\theta)^2\nonumber\\
 &=\left(\frac{1+\cos^4\theta}{2}\right)^2.
 \label{eq:supp_bell_dominant_single_site_weight}
\end{align}
This factor controls the large-$N$ exponential scaling for
$0<\theta<\pi/2$. At $\theta=\pi/2$, additional terms share the same
leading factor and yield the same limiting density. Consequently,
\begin{align}
 \lim_{N\rightarrow\infty}
 \left[-\frac1N\log_2\overline{2^{-\Ba(\theta)}}\right]
 &=-2\log_2\left(\frac{1+\cos^4\theta}{2}\right)\nonumber\\
 &=2-2\log_2(1+\cos^4\theta).
 \label{eq:supp_random_phase_ba_density}
\end{align}
Both moment-averaged densities are exact within the independent-phase
ensemble. Their finite-size corrections differ, so the Bell curves cannot
be obtained by doubling the finite-size $\mathcal M_2$ curves.
Fig.~\ref{fig:supp_theory_magic_ordering}(b) compares the finite-size Bell
saturated values and their thermodynamic limit with the initial density in
Eq.~\eqref{eq:supp_initial_ba} and the Anderson phase-averaged density in
Eq.~\eqref{eq:supp_anderson_ba_phase_density}. The initial Bell density
decreases for $\theta>\pi/4$, while the thermodynamic interacting saturated value
increases, displaying the opposite angular orderings.

The physical consequence is the opposite of the initial ordering.  We first
differentiate the $\mathcal M_2$ density line by line:
\begin{align}
 \frac{\partial}{\partial\theta}
 [1-\log_2(1+\cos^4\theta)]
 &=-\frac{1}{\ln2}\frac{1}{1+\cos^4\theta}
 \frac{\partial\cos^4\theta}{\partial\theta}\nonumber\\
 &=\frac{4\sin\theta\cos^3\theta}
 {\ln2(1+\cos^4\theta)},\nonumber\\
 &\geq0.
 \label{eq:supp_late_m2_monotonicity}
\end{align}
For the Bell density, the same substitution gives an additional factor of two:
\begin{align}
 \frac{\partial}{\partial\theta}
 [2-2\log_2(1+\cos^4\theta)]
 &=-\frac{2}{\ln2}\frac{1}{1+\cos^4\theta}
 \frac{\partial\cos^4\theta}{\partial\theta}\nonumber\\
 &=\frac{8\sin\theta\cos^3\theta}
 {\ln2(1+\cos^4\theta)},\nonumber\\
 &\geq0.
 \label{eq:supp_late_ba_monotonicity}
\end{align}
Here we used
$\partial_\theta\cos^4\theta=-4\sin\theta\cos^3\theta$.  Both late-time
thermodynamic derivatives are therefore nonnegative, whereas the initial derivatives in
Eqs.~\eqref{eq:supp_initial_m2_monotonicity} and
\eqref{eq:supp_initial_ba_monotonicity} are nonpositive.  Choose two tilts in
the stated interval with $\theta_1<\theta_2$.  Substituting the angles into the
initial and phase-averaged thermodynamic formulas gives
\begin{align}
 \lim_{N\rightarrow\infty}\frac{\mathcal M_2(0,\theta_2)}{N}
 &<\lim_{N\rightarrow\infty}\frac{\mathcal M_2(0,\theta_1)}{N},\nonumber\\
 \lim_{N\rightarrow\infty}
 \frac{\mathcal M_2^{\mathrm{sat}}(\theta_2)}{N}
 &>\lim_{N\rightarrow\infty}
 \frac{\mathcal M_2^{\mathrm{sat}}(\theta_1)}{N},\nonumber\\
 \lim_{N\rightarrow\infty}\frac{\Ba(0,\theta_2)}{N}
 &<\lim_{N\rightarrow\infty}\frac{\Ba(0,\theta_1)}{N},\nonumber\\
 \lim_{N\rightarrow\infty}
 \left[-\frac1N\log_2\overline{2^{-\Ba(\theta_2)}}\right]
 &>\lim_{N\rightarrow\infty}
 \left[-\frac1N\log_2\overline{2^{-\Ba(\theta_1)}}\right].
 \label{eq:supp_initial_late_reversal}
\end{align}
Eq.~\eqref{eq:supp_random_phase_i4_exact} confirms this ordering for
the finite sizes and angles in Fig.~\ref{fig:main_overview}(b).
If coarse-grained dynamics approaches these saturated values, continuity
guarantees a crossing. The thermodynamic saturated curve also lies above
the initial curve on this branch; their opposite orderings place the
larger-tilt state farther from its own saturated value. Both endpoints
are pattern independent: the longitudinal signs $\eta_i$ disappear from
the initial even moments and from the phase-averaged sums under
$z_i\mapsto\eta_i z_i$. Saturation-time ordering requires the separate
numerical analysis in Sec.~\ref{sec:supp_numerics} and End Matter
Fig.~\ref{fig:endmatter_localized_steady_state_time}, with localization
diagnosed independently through level statistics.

\begin{figure}[H]
  \centering
  \includegraphics[width=0.64\textwidth]{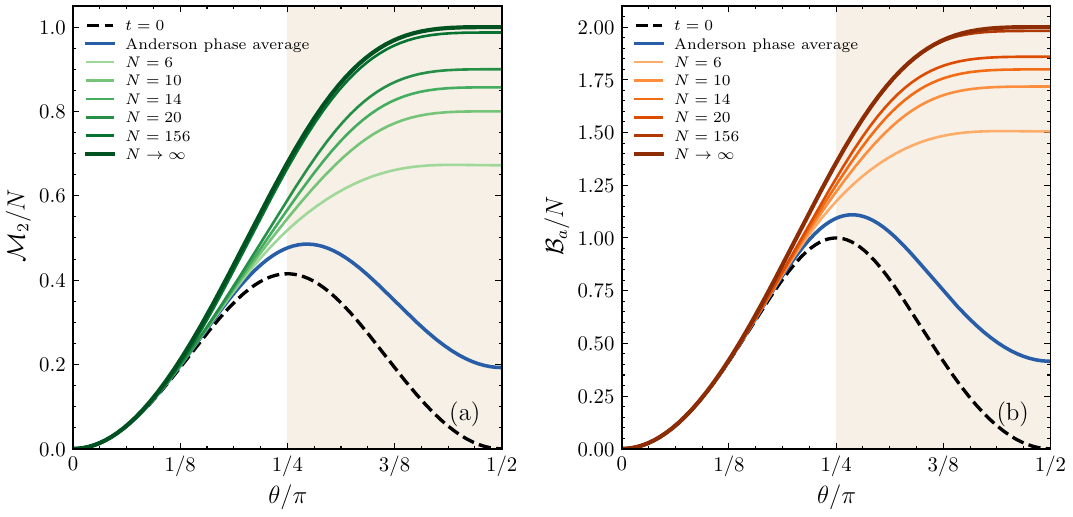}
  \caption[Analytical ordering of 2-SRE and additive Bell magic.]{\textit{Analytical
  ordering of 2-SRE and additive Bell magic.}
  (a) $\mathcal M_2/N$: the black dashed curve is the initial density
  [Eq.~\eqref{eq:supp_initial_m2}], and the blue curve is the Anderson
  local-phase average. Green curves show the interacting random-phase
  saturated values at $N=6,10,14,20,156$ and $N\to\infty$, from light to dark
  [Eqs.~\eqref{eq:supp_random_phase_i4_exact} and
  \eqref{eq:supp_random_phase_m2_density}].
  (b) $\Ba/N$: the black dashed and blue curves are the initial and
  Anderson phase-averaged densities
  [Eqs.~\eqref{eq:supp_initial_ba} and
  \eqref{eq:supp_anderson_ba_phase_density}]. Orange curves show the interacting random-phase saturated values
  at the same sizes $N=6,10,14,20,156$ and $N\to\infty$, from light
  to dark. The finite-size curves retain all contributions to the
  independent-phase average of Eq.~\eqref{eq:supp_bell_eighth_moment_expanded};
  the thermodynamic limit is Eq.~\eqref{eq:supp_random_phase_ba_density}.
  Shading marks $\theta>\pi/4$ in both panels. Phase-averaged densities
  are obtained by averaging $\mathcal I_4$ in (a) and $2^{-\Ba}$ in (b)
  before taking $-N^{-1}\log_2$. Numerical saturation times are compared
  in End Matter Fig.~\ref{fig:endmatter_localized_steady_state_time}.}
  \label{fig:supp_theory_magic_ordering}
\end{figure}
\par
\endgroup

\section{Numerical analysis of magic Mpemba dynamics across localized models}
\label{sec:supp_numerics}

\subsection{Exact evaluation of the 2-SRE with the XOR--FWHT algorithm}
\label{sec:supp_numerical_methods}

The $N=14$ trajectories underlying End Matter
Fig.~\ref{fig:endmatter_localized_steady_state_time} require exact magic
evaluation across a long logarithmic time grid. We use the
XOR--FWHT algorithm~\cite{SI-Huang2025XORFWHT}, which combines bitwise XOR
with a fast Walsh--Hadamard transform (FWHT), avoiding Pauli-string
sampling. Its working formulas are given below.

For a normalized state
$\ket{\psi}=\sum_{\bm x\in\mathbb Z_2^N}\psi_{\bm x}\ket{\bm x}$, let
$d=2^N$.  The indices $\bm x,\bm u,\bm v$ are $N$-bit strings, with
$z_i=(-1)^{x_i}$, $\oplus$ denoting bitwise XOR, and dot products evaluated
modulo two.  Up to an irrelevant phase, each Pauli string is
$X^{\bm u}Z^{\bm v}$.  For a fixed shift $\bm u$, form the length-$d$ array
$g_{\bm u}$ and apply the \emph{unnormalized} FWHT:
\begin{align}
 g_{\bm u}(\bm x)&=\psi^*_{\bm x\oplus\bm u}\psi_{\bm x},
 \label{eq:supp_fwht_xor_correlations}\\
 \widehat g_{\bm u}(\bm v)
 &=[\operatorname{FWHT}g_{\bm u}](\bm v)
 =\sum_{\bm x}(-1)^{\bm v\cdot\bm x}g_{\bm u}(\bm x)
 =\bra{\psi}X^{\bm u}Z^{\bm v}\ket{\psi}.
 \label{eq:supp_fwht_pauli_transform}
\end{align}
Thus one transform gives all $d$ expectation values sharing the same
$X$-mask.  Taking absolute fourth powers removes the Pauli phases, so
Eq.~\eqref{eq:supp_m2_moment} gives
\begin{equation}
 \mathcal M_2(\psi)
 =-\log_2\Ifour(\psi),\qquad
 \Ifour(\psi)=\frac1d\sum_{\bm u,\bm v}
 \left|\widehat g_{\bm u}(\bm v)\right|^4.
 \label{eq:supp_fwht_final_m2}
\end{equation}
The factor $1/d$ applies to the unnormalized transform defined above.
Algorithm~\ref{alg:supp_xor_fwht} summarizes the evaluation.

\begingroup
\algrenewcommand{\algorithmicrequire}{\textbf{Input:}}
\algrenewcommand{\algorithmicensure}{\textbf{Output:}}
\begin{algorithm}[H]
\caption{XOR--FWHT evaluation of $\mathcal M_2$}
\label{alg:supp_xor_fwht}
\begin{algorithmic}[1]
\Require Normalized amplitudes $\{\psi_{\bm x}\}_{\bm x\in\mathbb Z_2^N}$,
         with $d=2^N$.
\Ensure $\mathcal M_2(\psi)$.
\State $\Ifour\gets0$
\For{each $\bm u\in\mathbb Z_2^N$}
  \State \hspace*{\algorithmicindent}Construct $g_{\bm u}(\bm x)\gets\psi^*_{\bm x\oplus\bm u}\psi_{\bm x}$
         for all $\bm x\in\mathbb Z_2^N$
  \State \hspace*{\algorithmicindent}$\widehat g_{\bm u}\gets\operatorname{FWHT}(g_{\bm u})$
         \Comment{Unnormalized transform}
  \State \hspace*{\algorithmicindent}$\Ifour\gets\Ifour+d^{-1}\sum_{\bm v}|\widehat g_{\bm u}(\bm v)|^4$
\EndFor
\State \Return $\mathcal M_2(\psi)=-\log_2\Ifour$
\end{algorithmic}
\end{algorithm}
\endgroup

Each FWHT consists of $N$ butterfly stages,
$(a,b)\mapsto(a+b,a-b)$, and costs $\mathcal O(Nd)$.  The $d$ shifts
therefore require $\mathcal O(N4^N)$ time; the sequential form in
Algorithm~\ref{alg:supp_xor_fwht} uses $\mathcal O(d)$ working memory.
The evaluation is exact up to floating-point roundoff.  We check its
normalization against direct Pauli enumeration for small $N$, the analytic
tilted-product-state value, and stabilizer states with $\mathcal M_2=0$.
The additive Bell magic $\Ba$ in the insets is evaluated from the same state
vectors using Eq.~\eqref{eq:supp_bell_eighth_moment}.

\subsection{Further numerical results on localized magic Mpemba dynamics}
\label{sec:supp_numerical_results}

The saturation-time comparison in End Matter
Fig.~\ref{fig:endmatter_localized_steady_state_time} uses three static
microscopic models and the effective \(\ell\)-bit model. Unlike the
kicked-chain processor trajectories, these exact numerical trajectories
resolve both saturated magic and the time required to approach it.
The microscopic models are open XXZ chains,
\begin{equation}
 H_{\mathrm{XXZ}}
 =-\sum_{i=0}^{N-2}(X_iX_{i+1}+Y_iY_{i+1})
 +\frac12\sum_{i=0}^{N-2}Z_iZ_{i+1}
 +\sum_{i=0}^{N-1}h_iZ_i.
 \label{eq:supp_numerical_xxz_hamiltonian}
\end{equation}
Only the longitudinal field differs among the three cases:
\begin{equation}
 h_i=
 \begin{cases}
 W\zeta_i,
 &\text{random disorder},\\[2pt]
 W\cos[2\pi\beta(i+1)+\varphi],
 &\text{quasiperiodic},\\[2pt]
 F(i+1)+\kappa(i+1)^2,
 &\text{Stark}.
 \end{cases}
 \label{eq:supp_numerical_fields}
\end{equation}
Here each \(\zeta_i\) is an independent random number drawn uniformly from the
interval \([-1,1]\), the quasiperiodic wavenumber is set by the golden ratio
\(\beta=(\sqrt5-1)/2\), and the Stark curvature is
\(\kappa=0.075\).  Spin labels run from $0$ to $N-1$; the factors $i+1$
specify the spatial origin of these static potentials.  The random and quasiperiodic results are averaged over
independent field realizations and phases \(\varphi\), respectively.  The weak/strong parameter
pairs used in the dynamics are \(W=2,10\) for the random and quasiperiodic
chains and \(F=0.5,6\) for the Stark chain.  The weak case is labeled
ergodic and the strong case localized in the figures.  The small
quadratic Stark curvature removes the extensive degeneracies of a perfectly
linear tilt while retaining the field-gradient localization mechanism.

For comparison with the dephasing mechanism derived in
Sec.~\ref{sec:supp_lbit_theory}, we also evolve the three-body truncation of
$H_{\ell\text{-bit}}$ in Eq.~\eqref{eq:supp_lbit_theory_hamiltonian}.
Here \(h_i=10\zeta_i\) with the same uniform \(\zeta_i\in[-1,1]\), while
\(J_{ij}=J_{\mathrm{int}}e^{-(j-i)/\xi}\gamma_{ij}\) and
\(K_{ijk}=J_{\mathrm{int}}e^{-(k-i)/\xi}\gamma_{ijk}\), with independent
zero-mean, unit-variance Gaussian variables \(\gamma_{ij},\gamma_{ijk}\).
Thus $\eta_i$ is reserved for the initial-state signs and $\xi$ for the
localization length.  We use \(\xi=0.5\),
\(J_{\mathrm{int}}=1\), and one fixed realization.  This convention is the
same as in Fig.~\ref{fig:supp_lbit_magic_dynamics}.

For every realization, ED gives
\(H\ket{E_n}=E_n\ket{E_n}\).  The tilted initial states are those of
Eq.~\eqref{eq:supp_tilted_product_state}, with
\(\theta/\pi=0.35,0.50\) for the dynamics shown below, and are evolved spectrally as
\(\ket{\psi(t,\theta)}=\sum_n c_n(\theta)e^{-\ii E_nt}\ket{E_n}\) with
\(c_n(\theta)=\braket{E_n|\psi_0(\theta)}\).
Each requested time is obtained directly from this spectral sum;
there is no time-stepping error accumulated over the long logarithmic window.
Although \(H_{\mathrm{XXZ}}\) conserves total \(Z\) magnetization, the tilted
states span multiple charge sectors.  Their magic dynamics is therefore
computed in the full \(2^N\)-dimensional Hilbert space.  The ergodic data are
displayed over \(0\leq t\leq20\), including the initial values, whereas
the localized and \(\ell\)-bit data use 53 logarithmically spaced times over
\(10^{-2}\leq t\leq10^{12}\).

We first diagnose localization independently of magic.  Since the level
statistics do conserve charge, we diagonalize the half-filling sector, order
its complete spectrum, and evaluate the adjacent gaps
\(\delta_n=E_{n+1}-E_n\) and their ratios
\(r_n=\min(\delta_n,\delta_{n+1})/\max(\delta_n,\delta_{n+1})\).

\begin{figure}[H]
  \centering
  \includegraphics[width=\textwidth]{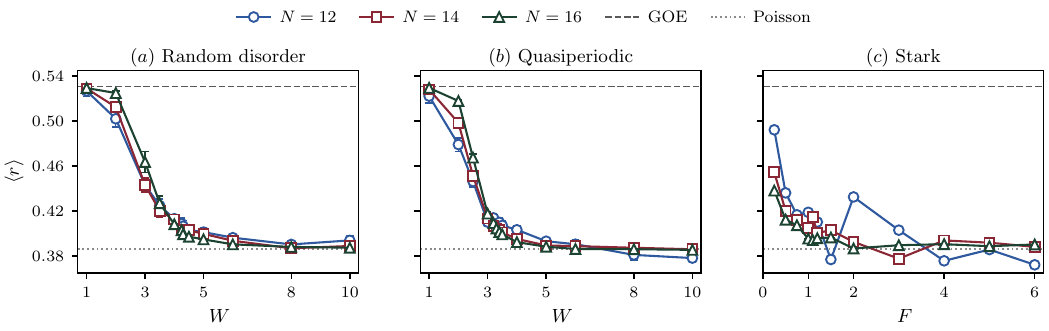}
  \caption[Level-spacing diagnostics for three localization mechanisms.]
  {\textit{Level-spacing diagnostics for three localization mechanisms.}
  Mean adjacent-gap ratio \(\langle r\rangle\) of the half-filling spectrum of
  the XXZ chain, against field strength, for (\(a\)) random disorder,
  (\(b\)) a quasiperiodic potential and (\(c\)) a Stark potential.  Colors
  label the chain length \(N=12,14,16\).  Error bars are the standard error of the mean (SEM) over ten
  disorder realizations or quasiperiodic phases; the Stark data are single
  deterministic samples.  The dashed and dotted horizontal lines mark the Gaussian orthogonal ensemble (GOE)
  and Poisson reference values, \(0.5307\) and \(0.3863\), respectively, so that the weak-
  and strong-field ends of each scan identify the ergodic and localized
  parameter sets used in Figs.~\ref{fig:supp_disorder_magic_dynamics}--\ref{fig:supp_stark_magic_dynamics}.}
  \label{fig:supp_r_ratio}
\end{figure}

Fig.~\ref{fig:supp_r_ratio} separates the regimes used below.  For random
disorder, the \(N=16\) mean changes from
\(\langle r\rangle=0.525\) at \(W=2\) to \(0.387\) at \(W=10\); for the
quasiperiodic chain it changes from \(0.518\) to \(0.386\).  The weak-field
values approach the GOE reference as \(N\) grows,
whereas the strong-field values approach the Poisson reference.  The Stark
scan has stronger finite-size oscillations because it is a single
deterministic potential, but its \(F=6\) values also lie near Poisson for the
larger sizes.  Thus the strong-field dynamics below is analyzed only after an
independent spectral localization diagnostic.

Figs.~\ref{fig:supp_disorder_magic_dynamics}--\ref{fig:supp_stark_magic_dynamics}
compare the two retained tilts at \(N=14\) in a common table layout.
The columns identify the TFS, TNS, and TDWS initial states through the same
schematics used in main-text Fig.~\ref{fig:main_kicked_l10}; the upper and
lower rows identify the localized and ergodic regimes, respectively, with
the field strength given beside each row.  Main panels show \(\mathcal M_2\),
and insets show \(\Ba\) for the same states and times.

\begin{figure}[H]
  \centering
  \includegraphics[width=\textwidth]{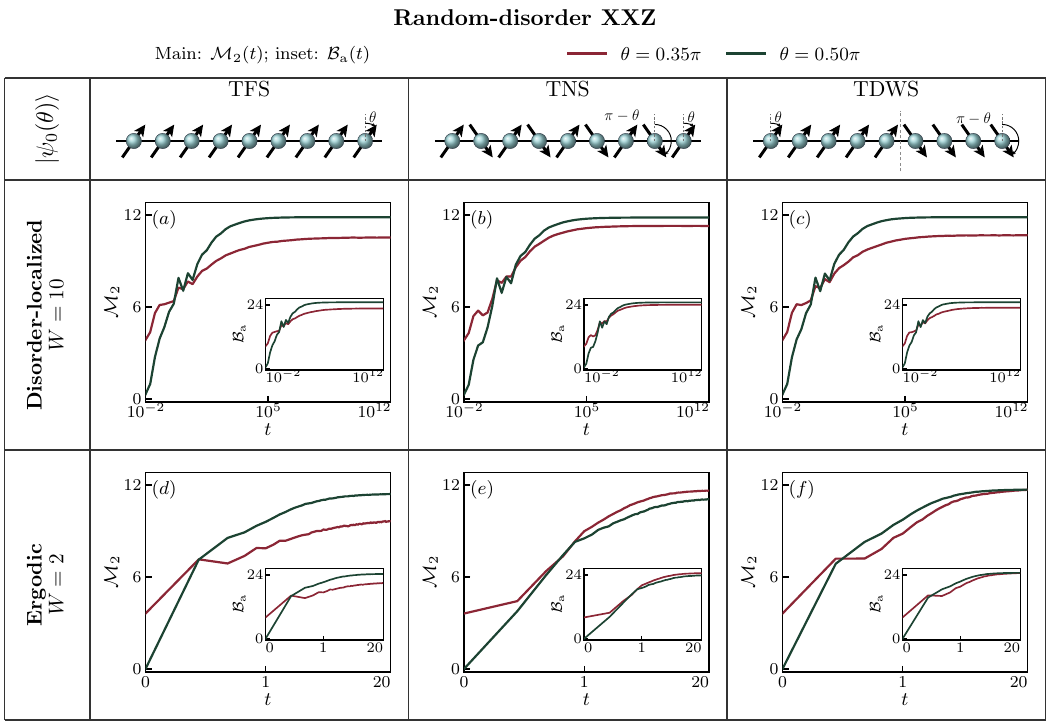}
  \caption[Random-disorder XXZ magic dynamics.]{\textit{Random-disorder XXZ
  magic dynamics.}  \(N=14\) results for \(\theta/\pi=0.35,0.50\), with
  \(\mathcal M_2(t)\) in the main panels and \(\Ba(t)\) in the insets.
  Columns are TFS, TNS, and TDWS; panels ($a$)--($c$) are disorder-localized
  at \(W=10\), and ($d$)--($f$) are ergodic at \(W=2\).
  Lines and bands give the mean and SEM over 50 disorder realizations.
  Localized time axes are logarithmic.  Ergodic axes retain \(t=0\)
  continuously, with a linear scale for \(0\leq t\leq0.2\) and a
  logarithmic scale thereafter up to \(t=20\).}
  \label{fig:supp_disorder_magic_dynamics}
\end{figure}

In the random-disorder chain, the localized \(0.50\pi\) curve overtakes
the \(0.35\pi\) curve for each initial-state family.  The weak-field response
is more pattern sensitive: TFS shows a clear reversal, the TDWS late-time
curves become close, and TNS does not exhibit the same sustained reversal
in its late-time values.  The localized comparison is therefore distinct
from the pattern-selective ergodic dynamics.

\begin{figure}[H]
  \centering
  \includegraphics[width=\textwidth]{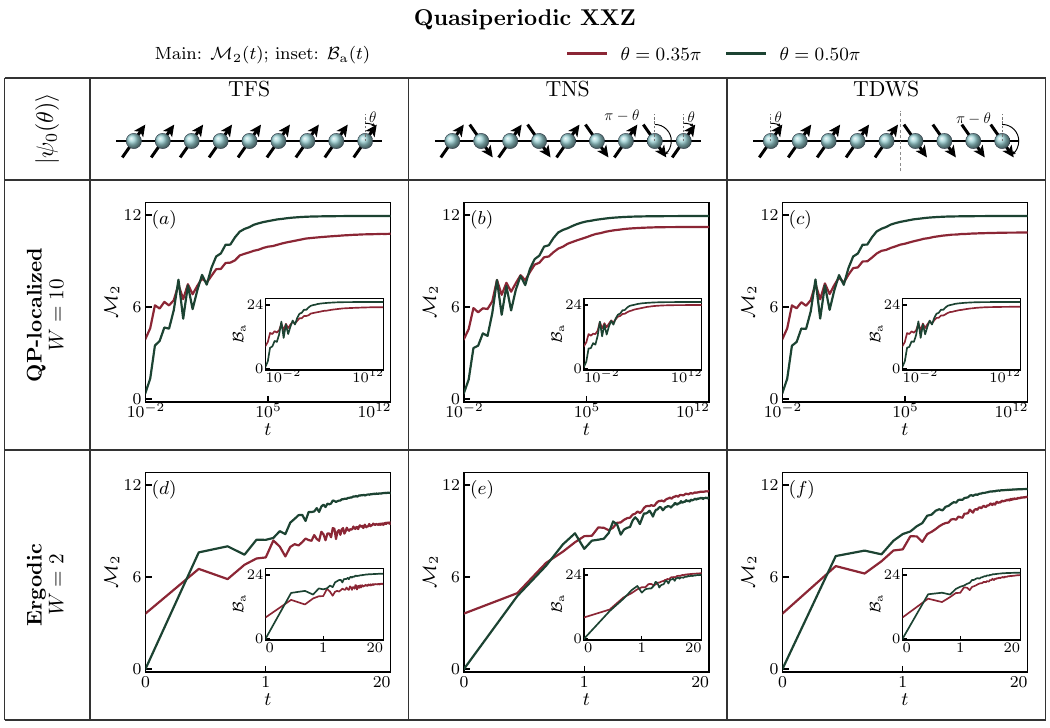}
  \caption[Quasiperiodic XXZ magic dynamics.]{\textit{Quasiperiodic XXZ magic
  dynamics.}  \(N=14\) results for \(\theta/\pi=0.35,0.50\), in the layout
  of Fig.~\ref{fig:supp_disorder_magic_dynamics}.
  Panels ($a$)--($c$) are quasiperiodic-localized at \(W=10\), and
  ($d$)--($f$) are ergodic at \(W=2\).
  Lines and bands give the mean and SEM over 50 quasiperiodic phases;
  main panels and insets show \(\mathcal M_2(t)\) and \(\Ba(t)\), respectively.}
  \label{fig:supp_qp_magic_dynamics}
\end{figure}

Replacing random disorder by a quasiperiodic field preserves the localized
reversal across all three initial-state families.  In the weak-field row,
TFS and TDWS show a clear reversal between the two retained tilts, whereas
TNS retains the opposite late-time ordering.  This contrast supports the
initial-state dependence observed in the main-text kicked-XXZ quantum
simulation [Fig.~\ref{fig:main_kicked_l10}], without relying on its
particular Floquet construction.

\begin{figure}[H]
  \centering
  \includegraphics[width=\textwidth]{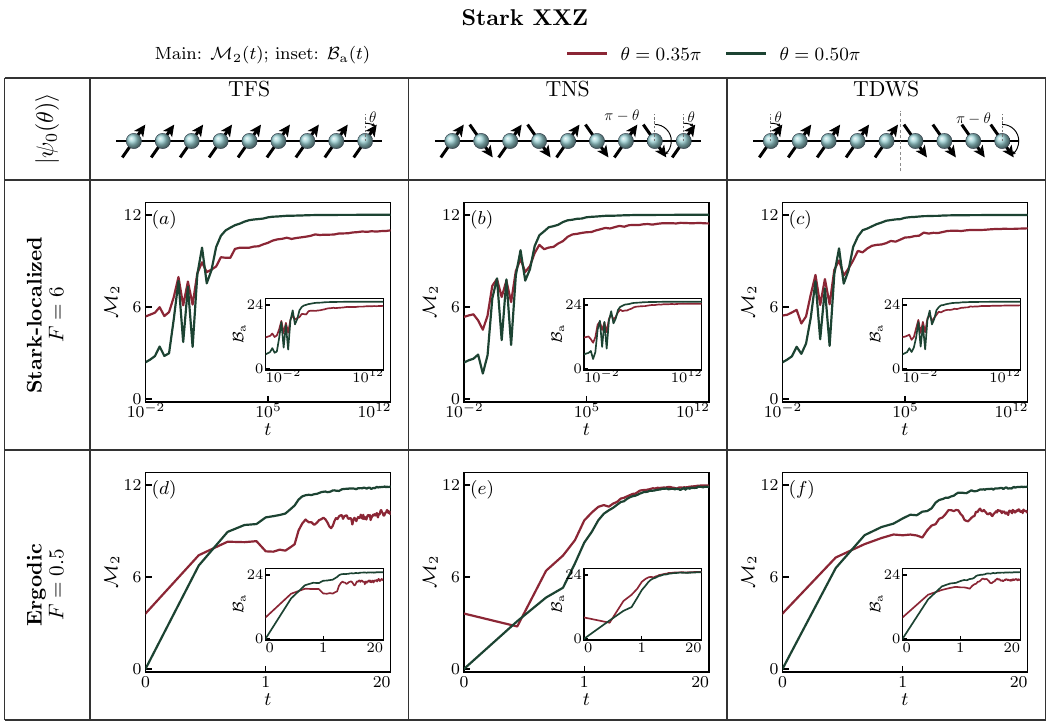}
  \caption[Stark XXZ magic dynamics.]{\textit{Stark XXZ magic dynamics.}
  \(N=14\) results for \(\theta/\pi=0.35,0.50\), in the layout of
  Fig.~\ref{fig:supp_disorder_magic_dynamics}.
  Panels ($a$)--($c$) are Stark-localized at \(F=6\), and ($d$)--($f$)
  are ergodic at \(F=0.5\), with curvature \(\kappa=0.075\).
  Main panels show \(\mathcal M_2(t)\) and insets \(\Ba(t)\).
  Each curve is a single deterministic realization, so no uncertainty band
  is shown.}
  \label{fig:supp_stark_magic_dynamics}
\end{figure}

The Stark chain exhibits the same contrast, despite its deterministic
potential and stronger transient oscillations.  TFS and TDWS reverse their
ordering in the weak-field row, whereas the TNS curves approach one another
without the same sustained late-time reversal.  In the strong-field row,
all three initial-state families reverse their ordering.

Across the three strong-field models, the common
tilted-product-state formula gives
\(\mathcal M_2(0,0.35\pi)>
\mathcal M_2(0,0.50\pi)\).  At long times, all three microscopic localization
mechanisms reverse this order to
\(\mathcal M_2^{\mathrm{sat}}(0.50\pi)>
\mathcal M_2^{\mathrm{sat}}(0.35\pi)\), for all three initial-state
families at \(N=14\).  The \(\Ba\) insets show the same ordering.
The localized reversal is therefore shared by TFS, TNS, and TDWS, although
their finite-size saturated values and transients need not coincide.
This common reversal is consistent with the explicit \(\ell\)-bit calculation in
Fig.~\ref{fig:supp_lbit_magic_dynamics}.

\clearpage
\begin{figure}[H]
  \centering
  \includegraphics[width=\textwidth]{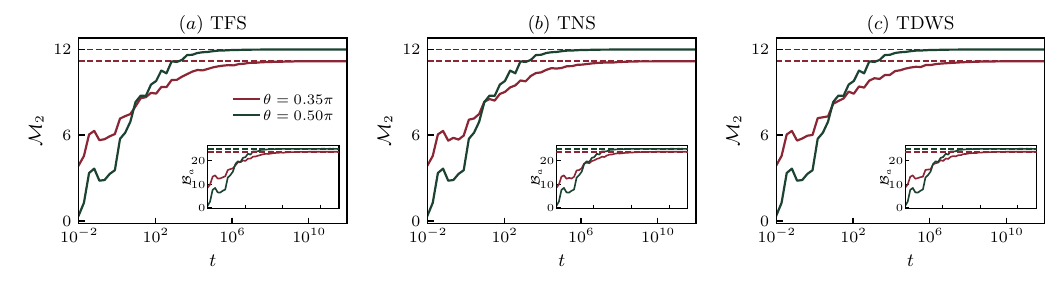}
  \caption[Exact interacting \(\ell\)-bit magic dynamics.]
  {\textit{Exact interacting \(\ell\)-bit magic dynamics.}
  \(\mathcal M_2(t)\) at \(N=14\) for ($a$) TFS, ($b$) TNS, and
  ($c$) TDWS, with \(\theta/\pi=0.35,0.50\).
  The more tilted state starts with less magic and reaches a higher saturated value.
  Main-panel dashed lines show the finite-size phase-averaged
  \(\mathcal M_2\) prediction from Eq.~\eqref{eq:supp_random_phase_i4_exact}.
  Insets show the additive Bell magic \(\Ba(t)\); their dashed lines are
  numerical estimates of the saturated values from the final five time points, not the
  thermodynamic approximation in Eq.~\eqref{eq:supp_random_phase_ba_density}.
  Both quantities exhibit the same magic-ordering reversal in the three-body
  model with \(\xi=0.5\).}
  \label{fig:supp_lbit_magic_dynamics}
\end{figure}

The additive Bell magic and the 2-SRE are distinct finite-size quantities,
so the matching reversals in Fig.~\ref{fig:supp_lbit_magic_dynamics}
provide an additional check beyond the fourth Pauli moment.  The
\(\mathcal B_a\) reference lines use numerical late-time values because
Eq.~\eqref{eq:supp_random_phase_ba_density} retains only the dominant
thermodynamic contraction.

The saturation times in End Matter
Fig.~\ref{fig:endmatter_localized_steady_state_time} use the normalized
distance defined in the End Matter,
\(\Delta_\theta(t)=|\mathcal M_2(t,\theta)-\mathcal M_2^{\mathrm{sat}}(\theta)|
/|\mathcal M_2(0,\theta)-\mathcal M_2^{\mathrm{sat}}(\theta)|\).
We estimate \(\mathcal M_2^{\mathrm{sat}}\) separately for each realization
from its final five logarithmic time points. The normalization gives
\(\Delta_\theta(0)=1\) and compares fractional progress toward each
state's own saturated value. For the random-disorder and quasiperiodic
models, we normalize each realization separately and average
\(\Delta_\theta(t)\) over 50 realizations before extracting
\(t_{\mathrm{ss}}\).

The saturation time \(t_{\mathrm{ss}}(\theta)\) is the earliest sampled
time from which \(\Delta_\theta(t)<\epsilon=0.1\) at every remaining
sample up to \(t_{\max}=10^{12}\). If the tolerance is not maintained
within this window, we report the lower bound
\(t_{\mathrm{ss}}>t_{\max}\). For all four localized models and the
three initial-state families,
\(t_{\mathrm{ss}}(0.50\pi)<t_{\mathrm{ss}}(0.35\pi)\).
Several of the additional \(\theta=0.20\pi\) trajectories reach the
tolerance only at very late times or remain outside it within the
simulated window. Varying the number of final time points used to
estimate saturation from three to seven leaves the ordering of
\(0.50\pi\) and \(0.35\pi\) unchanged. For the states studied, this saturation-time inversion
accompanies the reversal of saturated magic ordering.

For the sizes and states studied, random-field, quasiperiodic, and Stark
chains reproduce the reversed saturated-magic ordering of the effective
\(\ell\)-bit model across all three initial-state patterns. Their
saturation-time inversion [End Matter
Fig.~\ref{fig:endmatter_localized_steady_state_time}] extends the evidence
for dephasing-induced MME beyond the directly encoded model. The
level-statistics scan locates the chosen parameters in the ergodic and
localized regimes at these finite sizes; it does not establish a sharp
thermodynamic transition~\cite{SI-Sierant2025MBLReview}.

\clearpage
\begingroup
\makeatletter
\let\thebibliography\NAT@thebibliography
\let\endthebibliography\endNAT@thebibliography
\let\NAT@bibsetnum\NATx@bibsetnum
\makeatother
\renewcommand{\bibsection}{\section*{References}}
\arxivSIbibliographytrue
% One-file arXiv bibliography: main-text and SI lists retain separate numbering.
\ifarxivSIbibliography
\else
\fi

\endgroup

\clearpage
\listsupplementaryfigures
\listsupplementarytables

\endgroup
\end{document}